\pdfoutput=1
\documentclass[iop]{emulateapj}
\usepackage{graphicx}
\usepackage{amsmath,amssymb}
\usepackage{natbib}
\usepackage{mypack}
\usepackage[caption=false]{subfig}
\usepackage[breaklinks]{hyperref}
\usepackage{setspace}
\usepackage{tabularx}

\def\bfref{}

\begin{document}

\title{The Green Bank Northern Celestial Cap Pulsar Survey - I: Survey Description, Data Analysis, and Initial Results}
\author{K.~Stovall$^{1,2,3}$, R.~S.~Lynch$^{4}$, S.~M.~Ransom$^{5}$, A.~M.~Archibald$^{4,6}$, S.~Banaszak$^{7}$,
C.~M.~Biwer$^{7,8}$, J.~Boyles$^{9}$, L.~P.~Dartez$^{1}$, D.~Day$^{7}$, A.~J.~Ford$^{1}$, J.~Flanigan$^{7}$, A.~Garcia$^{1}$,
J.~W.~T.~Hessels$^{6,10}$, J.~Hinojosa$^{1}$, F.~A.~Jenet$^{1}$, D.~L.~Kaplan$^{7,11}$, C.~Karako-Argaman$^{4}$, V.~M.~Kaspi$^{4}$,
V.~I.~Kondratiev$^{6,12}$, S.~Leake$^{1}$, D.~R.~Lorimer$^{13}$, G.~Lunsford$^{1}$, J.~G.~Martinez$^{1}$, A.~Mata$^{1}$,
M.~A.~McLaughlin$^{13}$, M.~S.~E.~Roberts$^{14}$, M.~D.~Rohr$^{7}$, X.~Siemens$^{7}$, I.~H.~Stairs$^{15}$,
J.~van~Leeuwen$^{6}$, A.~N.~Walker$^{7}$, \& B.~L.~Wells$^{7,16}$}

\keywords{surveys -- pulsars: general -- pulsars: individual (PSRs J0214+5222, J0636+5129, J0645+5158, J1434+7257, J1816+4510) -- binaries: general -- binaries: eclipsing}

\footnotetext[1]{Center for Advanced Radio Astronomy, University of Texas at Brownsville, One West University Boulevard, Brownsville, Texas 78520, USA; stovall.kevin@gmail.com}
\footnotetext[2]{Department of Physics and Astronomy, University of Texas at San Antonio, San Antonio, TX, USA}
\footnotetext[3]{Department of Physics and Astronomy, University of New Mexico, Albuquerque, NM, USA}
\footnotetext[4]{Department of Physics, McGill University, 3600 University Street, Montreal, QC H3A 2T8, Canada}
\footnotetext[5]{National Radio Astronomy Observatory, 520 Edgemont Road, Charlottesville, VA 22901, USA}
\footnotetext[6]{ASTRON, the Netherlands Institute for Radio Astronomy, Postbus 2, 7990 AA, Dwingeloo, The Netherlands}
\footnotetext[7]{Physics Dept., U. of Wisconsin - Milwaukee, Milwaukee WI 53211, USA}
\footnotetext[8]{Department of Physics, Syracuse University, Syracuse NY 13244, USA}
\footnotetext[9]{Department of Physics and Astronomy, Western Kentucky University, Bowling Green, KY 42101 USA}
\footnotetext[10]{Astronomical Institute `Anton Pannekoek', University of Amsterdam, Postbus 94249, 1090 GE Amsterdam, The Netherlands}
\footnotetext[11]{Department of Astronomy, University of Wisconsin-Madison, 475 North Charter Street, Madison, Wisconsin 53706-1582, USA}
\footnotetext[12]{Space Center of the Lebedev Physical Institute, Profsoyuznaya str. 84/32, Moscow 117997, Russia}
\footnotetext[13]{Department of Physics, West Virginia University, 210E Hodges Hall, Morgantown, WV 26506, USA}
\footnotetext[14]{Eureka Scientiﬁc, Inc., 2452 Delmer Street, Suite 100, Oakland, CA 94602-3017, USA}
\footnotetext[15]{Department of Physics and Astronomy, University of British Columbia, 6224 Agricultural Road, Vancouver, BC V6T 1Z1, Canada}
\footnotetext[16]{Department of Atmospheric Science, Colorado Sate University, Fort Collins, CO}

\begin{abstract}
We describe an ongoing search for pulsars and {\bfref dispersed pulses of radio emission, such as those from
rotating radio transients (RRATs) and fast radio bursts (FRBs),} at 350\,MHz using the Green Bank
Telescope.  With the Green Bank Ultimate Pulsar Processing Instrument, we record 100\,MHz of bandwidth divided
into 4,096 channels every 81.92 $\mu s$. This survey will cover the entire sky
visible to the Green Bank Telescope ({\bfref $\delta > -40\degr$, or 82\% of the sky}) {\bfref and outside of the
Galactic Plane will be sensitive enough to detect slow pulsars and low dispersion measure ($<$30 $\mathrm{pc\,cm^{-3}}$)
millisecond pulsars (MSPs) with a 0.08 duty cycle down to 1.1 mJy. For pulsars with a spectral index of $-$1.6, we
will be 2.5 times more sensitive than previous and ongoing surveys over much of our survey region.}
Here we describe the survey, the data analysis pipeline, initial discovery parameters for 62
pulsars, and timing solutions for 5 new pulsars. PSR J0214$+$5222 is an MSP
in a long-period (512 days) orbit {\bfref and has an optical counterpart identified in archival
data}. PSR J0636$+$5129 is an MSP in a very short-period (96 minutes) orbit {\bfref with a very low mass
companion (8 $M_\mathrm{J}$)}. PSR J0645$+$5158 is an isolated MSP {\bfref with a timing residual
RMS of 500 ns and } has been added to pulsar timing array experiments. PSR
J1434$+$7257 is an isolated, intermediate-period pulsar that has been partially recycled. PSR J1816$+$4510 is
an eclipsing MSP in a short-period orbit (8.7 hours) {\bfref and may have recently completed its spin-up phase}.
\end{abstract}
\maketitle

\section{Introduction}\label{sec:gbnccintro}
Since their initial discovery over 45 years ago~\citep{1968Natur.217..709H}, radio pulsars have
been excellent laboratories for a wide range of physics. They have provided some of the most
stringent tests of general relativity~\citep{1989ApJ...345..434T,2006Sci...314...97K}, as well
as alternative theories of gravity~\citep{2012MNRAS.423.3328F}. Some pulsars have been identified
as laboratories for extreme physics due to their strong magnetic fields~\citep{2008Sci...319.1802G}
or exceptionally large densities~\citep{2010Natur.467.1081D,2013Sci...340..448A}. Pulsars are
important probes of the Galactic neutron star population; those in binary (or multi-body) systems also
provide probes of their companion(s), which can be, e.g., planets~\citep{1992Natur.355..145W},
a white dwarf~\citep[e.g.][]{2005ASPC..328..357V}, a main sequence star~\citep{1992ApJ...387L..37J,1994ApJ...423L..43K},
another neutron star~\citep{1975ApJ...195L..51H}, or even another pulsar~\citep{2003Natur.426..531B}. Despite the discovery of
over 2,300 pulsars, pulsar searches are still uncovering novel systems,
including extremely fast-spinning millisecond pulsars~\citep[MSPs;][]{2006Sci...311.1901H}, MSPs in orbit
with main sequence star companions~\citep{2008Sci...320.1309C}, a very dense planetary
mass companion~\citep{2011Sci...333.1717B}, pulsars which appear to be transitionary
objects from low mass X-ray binaries to MSPs~\citep{2009Sci...324.1411A,2012ApJ...753..174K,2013Natur.501..517P},
and a hierarchical stellar triple system~\citep{2014Natur.505..520R}.

Radio pulsars, particularly MSPs, are excellent clocks spread throughout our galaxy. 
In recent years, there has been a strong effort to use an ensemble of MSPs, known as a pulsar
timing array (PTA), for direct gravitational wave~\citep[GW;][]{2011MNRAS.414.3117V,2012arXiv1210.6130M,2013ApJ...762...94D}
detection. Currently,
there are four PTA experiments ongoing, including the North American Nanohertz Observatory
for Gravitational Waves (NANOGrav), the Parkes Pulsar Timing Array, the European Pulsar
Timing Array, and a combination of these three known as the International Pulsar Timing Array.
A successful detection of GWs by a PTA will require a large number (20$-$40) of MSPs
spread across the sky, each of which must be timed with enough precision to predict its pulse
times-of-arrival to within tens to hundreds of nanoseconds. Recent pulsar surveys have shown that
we're still far from having found all of the best pulsar clocks in the Galaxy; thus pulsar surveys
are an important component in the effort to directly detect nano-Hertz gravitational waves.
Discovering MSPs that are suitable for PTAs is the primary driver for the survey we present here.

Two, complementary search strategies are employed to discover new pulsars. One is through targeted
searches of specific objects including supernova remnants (for a review, see~\citealt{2003ASPC..302..145C}),
globular clusters~\citep{1991Natur.352..219M,1994MNRAS.267..125B,2001ApJ...548L.171D,2005Sci...307..892R,2007ApJ...670..363H},
or high-energy counterparts, such as gamma-ray point sources identified by the \textit{Fermi Gamma-Ray
Space Telescope}~\citep[e.g.][]{2011AIPC.1357...40H,2011ApJ...727L..16R,2011MNRAS.414.1292K,2012ApJ...748L...2K}.
Such targeted searches can be highly efficient, however they are biased to what type of pulsar they can discover.
Searches of globular clusters yield large numbers
of MSPs, but they are typically very distant and therefore weak and strongly affected by propagation through the
ionized interstellar medium. \textit{Fermi}-identified point sources that are revealed to be pulsars are often MSPs (at
least off the Galactic plane), but in a large
fraction of these systems the companion is being ablated by the MSP. These systems often exhibit eclipses
for a significant fraction of the orbital period and have orbital effects which are detrimental to their
use in GW detection. Though there are certainly some \textit{Fermi}-discovered MSPs which are promising
for PTAs, there are also many important PTA sources which have no \textit{Fermi}
counterpart. The other approach to pulsar searches involves more shallow surveys of large areas of sky.  Such
surveys complement the targeted searches by discovering nearby sources (ideal for multi-wavelength follow-up,
and often good for PTAs) and filling out the total population.

Over the last five or so years, many new members of two related binary {\bfref pulsar} populations have been
discovered. {\bfref One population was first recognized with the discovery of the so-called ``black widow''
system~\citep[B1957$+$20;][]{1988Natur.333..237F}. It was called the black widow system due to evidence that
the companion is being ablated by the pulsar's wind. Roughly a dozen systems with similar characteristics, including
eclipses of the pulsar or DM variations of the pulsar's signal near superior conjunction, orbital periods ranging from about
$1.5-30$ hours, and very-low-mass ($M\sim0.01-0.05\,M_\mathrm{\odot}$) companions have been discovered
since. Pulsar systems with similar
characteristics to the black widow systems, but with low to moderate
mass companions ($M\sim0.15-0.7\,M_\mathrm{\odot}$), have also been discovered~\citep[e.g.][]{2009Sci...324.1411A}.
Given the similarities between these and the black widow systems, they are often referred to
as ``redback'' systems~\citep{2011AIPC.1357..127R}}.

\begin{figure}[h!]
\centering
\includegraphics[width=0.5\textwidth]{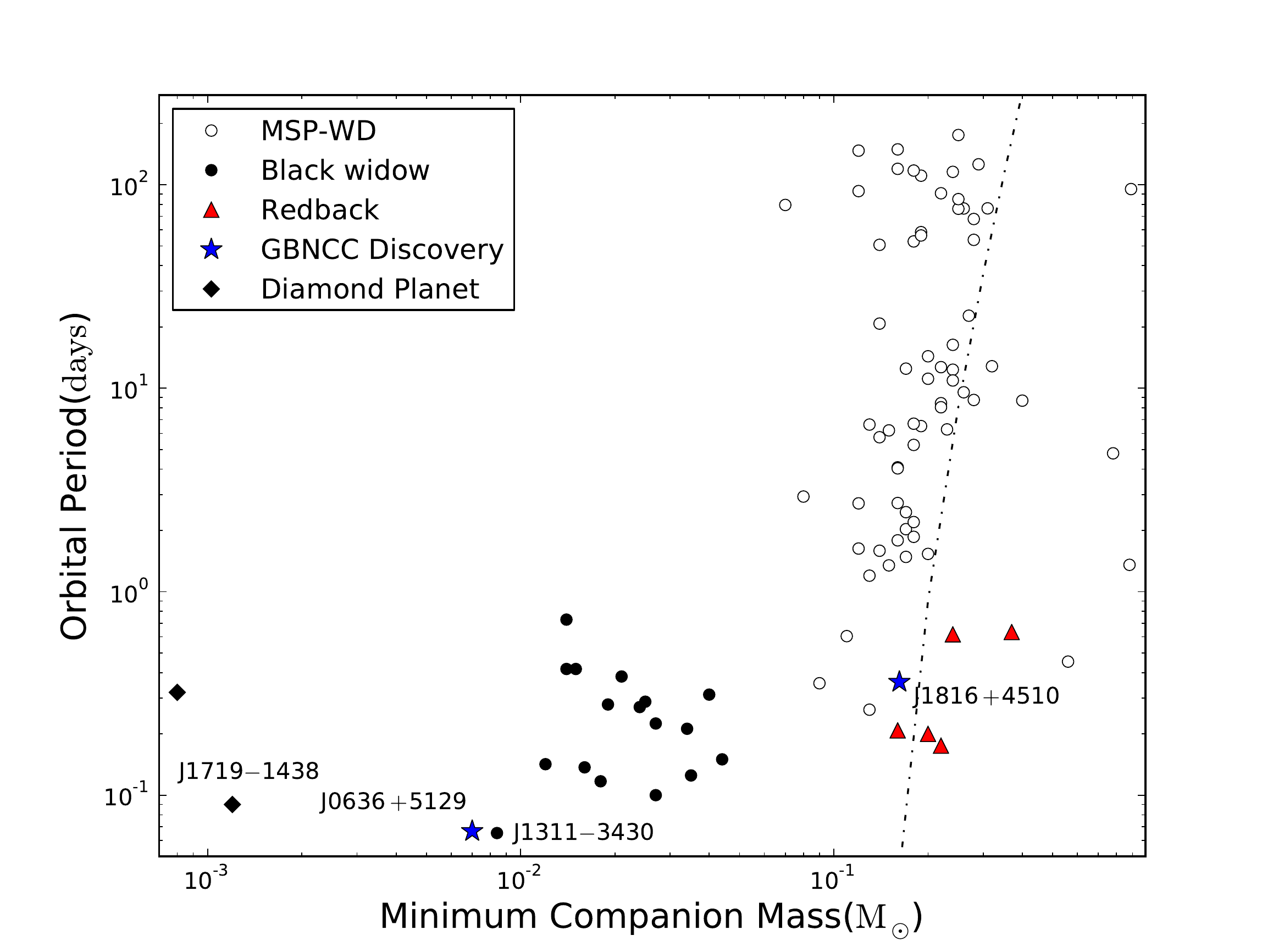}
\caption[Mass-Orbital Period Diagram for GBNCC Binary MSPs]{Minimum companion mass versus orbital period
for Galactic field MSPs ($P<10$ ms). Pulsars from the ATNF pulsar catalogue are shown as unfilled black
circles, known redbacks are filled red circles, known black widows are filled black circles, PSR J1719$-$1438
and another similar system described in~\cite{2013IAUS..291...53N} are
shown as black diamonds; and our discoveries are marked with blue stars. Parameters for redback and black widow
pulsars were taken from~\cite{2011AIPC.1357..127R} and Roberts, personal communication. The dashed line shows
the relation between companion mass and orbital period given by~\cite{ts99}.}
\label{fig:gbncc_massvpb}
\end{figure}

Another recently discovered population, the ``diamond planet'' systems, currently only contains two samples,
J1719$-$1438~\citep{2011Sci...333.1717B} and another system described
in~\cite{2013IAUS..291...53N}. The J1719{\bfref$-$1438} system contains a planetary mass companion
($1.2 M_\mathrm{J}$) in a very tight orbit ($P_\mathrm{orb}$=2.2~h). It is believed that these systems
resulted from an ultracompact X-ray binary system in which the companion narrowly avoided being completely
destroyed. We introduce these populations here since we will describe newly discovered sources in
Section~\ref{sec:gbnccdisc} which may be members of these populations. A plot showing the relationship
among typical binary pulsars with white dwarf companions, the black widow population, the
redback population and the diamond planet system is shown in Figure~\ref{fig:gbncc_massvpb}.

Driven by discovering exotic pulsar systems and the need for many more
PTA-worthy pulsars at a variety of angular separations from one another, there are several large-scale
pulsar surveys currently underway. Here we describe the Green Bank Northern Celestial Cap (GBNCC) survey.
The GBNCC will survey the entire sky visible to the Robert C. Byrd Green
Bank Radio Telescope (GBT) at 350 MHz for radio pulsars and fast radio transients. 
As of 2014 March 18, we have discovered 67 new radio pulsars{\bfref \footnote{For an up-to-date total, see
\url{http://arcc.phys.utb.edu/gbncc/}.}}, including 9 MSPs~($P$~$<$~10~ms) and 6
intermediate-spin-period pulsars~(10~ms~$<$~$P$~$<$~100~ms)
{\bfref after completing the processing of 20\% of the entire survey}
. We have also discovered an additional
7 rotating radio transients{\bfref \footnote{As of 2014 March 18, see footnote 1 to get an up-to-date total.}} (RRATs),
which will be presented in subsequent papers. In Section~\ref{sec:gbnccdesc},
we give a detailed description of the survey, followed by a description of the data analysis
and candidate rating in Section~\ref{sec:gbnccdata}. Section~\ref{sec:gbnccredet} lists re-detections of
previously known pulsars, which we use to estimate the sensitivity of our survey and provide for calibrating population simulations.
Section~\ref{sec:timing} describes the procedure we are
using to follow-up our discovered pulsars. In Section~\ref{sec:gbnccdisc},
we present results for 67 discovered pulsars; we give discovery parameters for
62 pulsars and timing solutions for 5 of our first discoveries. We are currently following up the{\bfref 5 pulsars presented here and our} other
discoveries (including the RRATs) using the GBT and the Low Frequency
Array~\citep[LOFAR;][]{2013A&A...556A...2V,2011A&A...530A..80S}.

\section{Survey Description}\label{sec:gbnccdesc}
\begin{figure}[h!]
\centering
\includegraphics[width=0.5\textwidth]{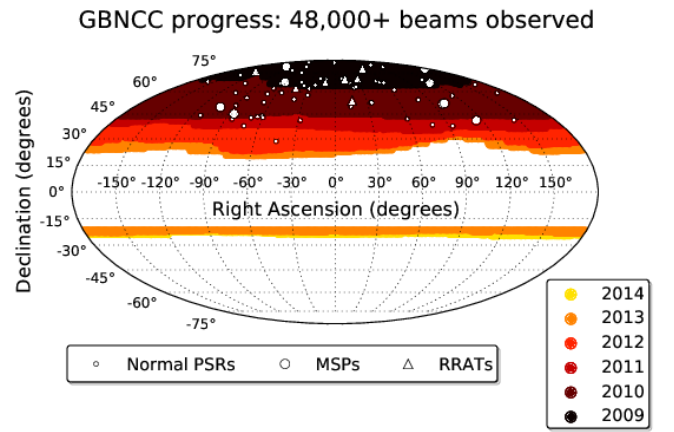}
\caption[GBNCC Coverage]{Sky covered to date by the GBNCC survey. Colors indicate
our coverage for each year from 2009 to 2014. Our discoveries are also shown; small
dots indicate normal pulsars, large circles correspond to MSPs, and triangles indicate
RRATs.
}
\label{fig:gbncc_skycoverage}
\end{figure}

After the success of the GBT 350 MHz Drift Scan survey~\citep[GBT350][]{2013ApJ...763...80B,2013ApJ...763...81L} as
well as with the installation of the new Green Bank Ultimate Pulsar
Processor\footnote{\url{https://safe.nrao.edu/wiki/bin/view/CICADA/GUPPiUsersGuide}} (GUPPI)
backend, we developed a plan to perform an all-sky pulsar survey{\bfref, known as the GBNCC survey,} using the GBT at 350 MHz.
{\bfref The GBNCC survey uses the GUPPI backend, which provides increased bandwidth and time resolution
over the Spigot~\citep{2005PASP..117..643K} backend used by the GBT350 survey.}
{\bfref The GBNCC} survey is divided into two stages: the first, which has been completed, covered the entire sky
north of $\delta=38^\circ$, and the second, which covers the remaining GBT visible sky,
is currently being carried out. {\bfref The positions
of our pointings were calculated from a generalized spiral set beginning at the North Pole as described
in~\cite{SaffKuijlaars97PointsOnSphere} using the Golden Section as the angle between successive points
for a set containing 152,000 points. The number of points used to cover the celestial sphere were chosen so
that the HWHM (18$'$) for the GBT's 350 MHz reciever would overlap with adjacent pointings by about 10\%
to provide good coverage of the entire sky.} Each pointing consists
of a 120-s integration on a particular sky location, with data recorded from 300 to 400 MHz
divided into 4,096 frequency channels. The vast majority of our pointings were recorded
with a sample time of 81.92 $\mu s$, but there were 207 pointings (all close to the northern celestial
pole) taken during the first few days of the survey with a sample time of 163.84 $\mu s$. 
Stage I of this survey was performed from 2009-10-23 to 2011-10-17 and resulted in $\approx30,000$ sky
pointings. {\bfref The data resulting from these pointings are still undergoing data processing. We have
processed about 80\% of these pointings and roughly 20\% of these candidates have been looked at by human
eye.} We have begun the second stage and {\bfref as of 2013 March 18 have} covered about 40\%
{\bfref (13,500 square degrees)} of the GBT visible sky.
Figure~\ref{fig:gbncc_skycoverage} shows our survey progress from 2009 until now.

The theoretical minimum detectable flux density for a pulsar survey is given by 
\begin{equation}
S_\mathrm{min}=\frac{\left(\mathrm{SNR_{min}}\right) T_\mathrm{sys}}{G \sqrt{n_\mathrm{p} t_\mathrm{int} \Delta f}} \sqrt{\frac{W}{P-W}}\mathrm{,}
\label{eqn:flux}
\end{equation}
where $\mathrm{SNR_{min}}$ is the minimum signal-to-noise ratio (SNR) required to detect a source, $T_\mathrm{sys}$ is
the system temperature in K, $G$ is the telescope gain in K/Jy, $n_\mathrm{p}$ is the number of polarizations summed,
$t_\mathrm{int}$ is the integration time in s, $\Delta f$ is the bandwidth in MHz, $W$ is the width of the pulse as
detected by the system in s, and $P$ is the pulsar's period {\bfref in s~\citep{1985ApJ...294L..25D}}. In our sensitivity
estimates, we used $\mathrm{SNR_{min}}$ values of either 12 or 15 to adjust for sensitivity differences
of surveys due to analog to digital conversion losses. We used a value of 12 for currently ongoing surveys
which use multi-bit systems and we used 15 for past surveys. The values of 12 and 15 are conservative
and account for sensitivity reduction due to radio frequency interference (RFI). {\bfref We note that although we
use an SNR of 12 for the estimated sensitivity of our survey, in the data analysis described in
Section~\ref{sec:gbnccdata}, we fold candidate pulsars down to a significance of 6 (the lowest
SNR of a pulsar blindly detected to date is about 8.5).} The GBT has a
gain of about 2 K/Jy at 350 MHz. The total system temperature is the receiver temperature ($T_\mathrm{rec}$)
plus the sky temperature ($T_\mathrm{sky}$). The receiver temperature at 350 MHz for the GBT is about 23 K.
The sky temperature is highly dependent on sky position and observing frequency, due to Galactic synchrotron
emission. Since we are estimating sensitivities for a large area of sky, we use the average
sky temperature of 34.4 K at 408 MHz~\citep{1982A&AS...47....1H} and calculate an estimate for $T_\mathrm{sky}$
at 350 MHz using the spectral index $-$2.6~\citep{1982A&AS...47....1H}. For the GBNCC survey, we use
$\Delta f$=80\,MHz to adjust for rolloff on the edge of the 100-MHz band, $t_\mathrm{int}$=120 s, and $n_\mathrm{p}$=2.
The pulse width ($W$) is not just a function of the pulsar, but also of incorrected pulse broadening due
to propagation through the interstellar medium. Dispersion in the interstellar medium widens the pulse as
\begin{equation}
\tau_\mathrm{DM}\bfref{\approx8.3}\,\mu s \times \left(\frac{\Delta f}{\mathrm{MHz}}\right)
\times \left(\frac{f}{\mathrm{GHz}}\right)^{-3}\times \left(\frac{\mathrm{DM}}{\mathrm{pc\,cm^{-3}}}\right)\mathrm{,}
\label{eqn:dispersion}
\end{equation}
where $f$ is the center frequency of a frequency channel and $\Delta f$ is the channel bandwidth. 
To minimize dispersive pulse broadening, we use 4,096 frequency channels across the observing bandwidth.
For dispersion measure (DM) = 100 $\mathrm{pc\,cm^{-3}}$, our setup would result in a dispersive pulse
smearing of $0.5$ ms. Another
effect of the interstellar medium is scattering, which broadens the pulse roughly according to
\begin{equation}
\log{\left(\frac{\tau_\mathrm{scatt}}{\mathrm{\mu s}}\right)}=-3.59+0.129 \log{\left(\mathrm{DM}\right)}+1.02 \log{\left(\mathrm{DM}\right)}^2-4.4 \log{\left(\frac{f}{\mathrm{GHz}}\right)}
\label{eqn:scattering}
\end{equation}
\citep{2002ASPC..278..227C}.
Since we are unable to mitigate the effect of scattering, it is difficult to
detect most MSPs {\bfref beyond} a DM of 100 $\mathrm{pc\,cm^{-3}}$ ($\tau_\mathrm{scatt}\sim0.8$ ms) at these frequencies.
We note that though the dispersion law in Equation~\ref{eqn:dispersion} is very accurate, the scattering
relation in Equation~\ref{eqn:scattering} is highly uncertain and the scattering for some lines-of-sight can
vary from this relation by over an order of magnitude. Scattering, combined with higher sky temperature, makes
the GBNCC survey less sensitive to distant pulsars in the Galactic plane, compared with higher-frequency surveys.
However, 74\% of our surveyed area is in directions with maximum expected DMs of less than 100
$\mathrm{pc\,cm^{-3}}$ for Galactic pulsars, as predicted by the NE2001 model.

\begin{figure}[h!]
\centering
\includegraphics[width=0.5\textwidth]{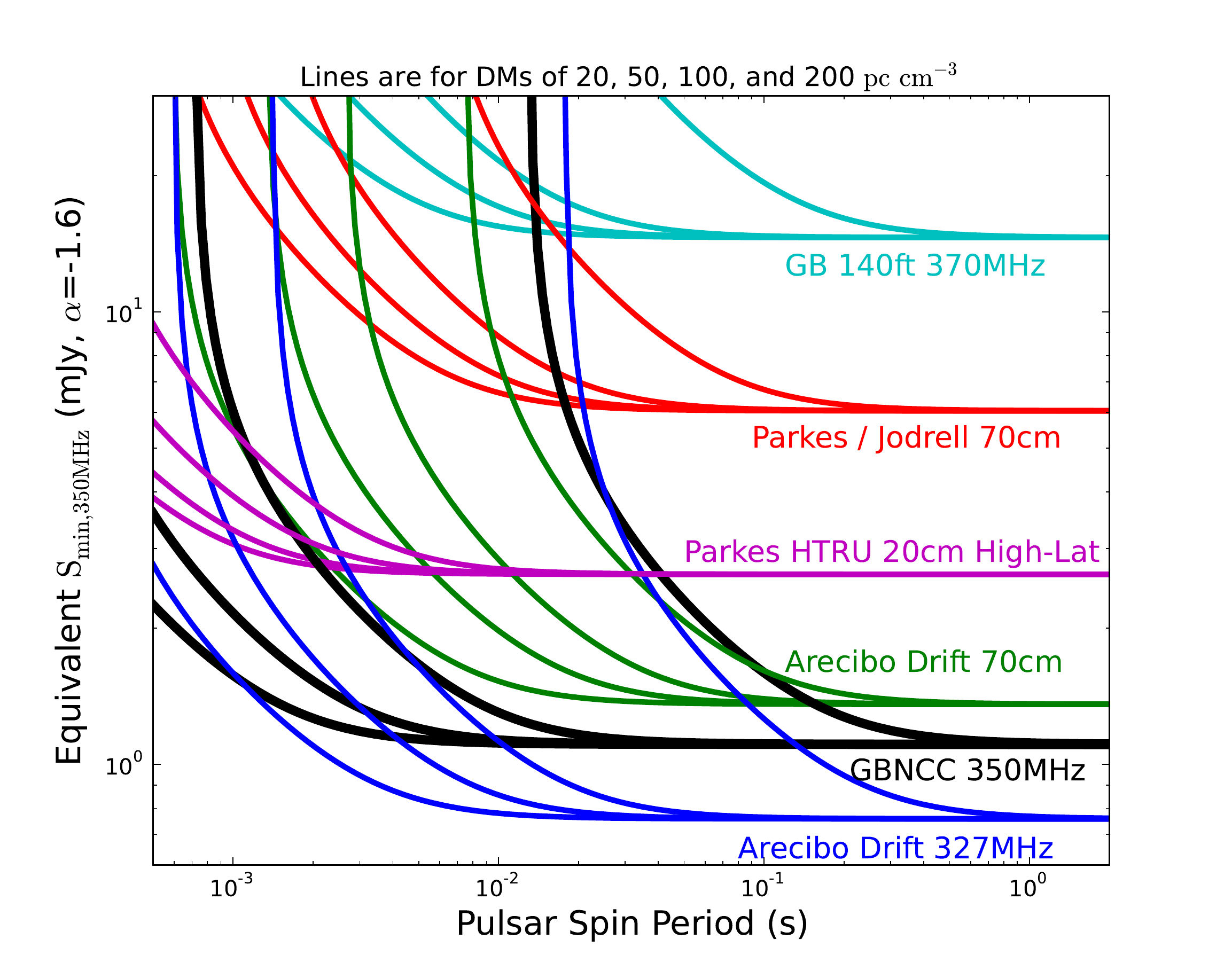}
\caption[GBNCC Estimated Sensitivity]{Estimated sensitivity of the GBNCC survey compared to the estimated
sensitivities of five other pulsar surveys. For each survey, sensitivities are calculated using DMs of
20, 50, 100, and 200~$\mathrm{pc\,cm^{-3}}$. In each case, the DM=20~$\mathrm{pc\,cm^{-3}}$ curve is
the leftmost and the DM=200~$\mathrm{pc\,cm^{-3}}$ curve is the rightmost. Sensitivities are scaled to
350 MHz using a spectral index of $-$1.6.}
\label{fig:sensitivity}
\end{figure}

Figure~\ref{fig:sensitivity} shows the estimated sensitivity of the GBNCC survey compared with five
past or ongoing surveys at DMs of 20, 50, 100, and 200 $\mathrm{pc\,cm^{-3}}$. The GBNCC survey is about
a factor of 2.5 more sensitive to low-DM pulsars in high Galactic latitude regions than the ongoing
High Time Resolution Universe (HTRU) surveys being performed with Parkes and
Effelsberg~\citep{2010MNRAS.409..619K}, is significantly more sensitive to MSPs than past Arecibo
surveys, and has comparable sensitivity for MSPs to the ongoing Arecibo Drift 327 MHz survey (AO327),
which covers the declination range $0^\circ>\delta>38^\circ$. We have not
compared our sensitivity to Galactic plane surveys, such as the Pulsar Arecibo L-Band Feed Array (PALFA)
survey~\citep{2006ApJ...637..446C,2013IAUS..291...35L}, the Parkes Multibeam Pulsar Survey (PMPS)~\citep{2001MNRAS.328...17M},
and the HTRU mid and low latitude surveys~\citep{2010MNRAS.409..619K} as these searches focus on low Galactic
latitudes, where we are not as sensitive due to higher $T_\mathrm{sky}$, dispersive smearing, and scattering - all of
which are much larger effects in our frequency range.

To estimate the number of millisecond pulsars detectable by the GBNCC, we have carried out simulations using
the snapshot population model described by~\cite{2013IAUS..291..237L}, which is normalized to match the results
of the Parkes Multibeam Pulsar Survey~\citep{2001MNRAS.328...17M}. Since this model provides pulsar luminosities
at 1.4 GHz, we have assumed an underlying radio spectral index distribution with a mean of $-1.4$ and unit standard
deviation, as found for the normal pulsar population by~\cite{2013MNRAS.431.1352B} to scale the luminosities to 350
MHz. Under these assumptions, the total number of MSPs expected to be detectable in Stage I of
the survey is $\sim 30$. The full survey is expected to detect of order 130 MSPs total. Future work will examine this
population in more detail to constrain the spectral index and luminosity distributions of MSPs.

\section{Data Analysis}\label{sec:gbnccdata}

For all-sky, high time/frequency resolution surveys such as the GBNCC, the data volume is very large. The entire
survey will result in {\bfref about 700 terabytes of raw} data. In addition to a large data volume,
the algorithms used to search the data require large amounts of computer time and
typically require high performance computing (HPC) clusters. The majority of the
data analysis for the GBNCC survey has been conducted using two HPC
systems. The Guillimin cluster operated by CLUMEQ and Compute Canada has accounted for
about 80\% of processing to date, while the Lonestar and Ranger Clusters operated by the Texas
Advanced Computing Center has accounted for the remaining 20\%. We have recently begun processing
data on a third HPC system at the University of Wisconsin-Milwaukee.

The data analysis process consists
of RFI excision, correction for the effects of interstellar dispersion, a search for periodic
signals, a single-pulse search, sifting of candidates, and then human inspection of diagnostic
plots created by the search algorithms. The following subsections describe each of these pieces
in more detail.

\subsection{RFI Excision}\label{subsec:rfi}
The data are first analyzed using the {\tt rfifind} tool from the
{\tt PRESTO}\footnote{\url{http://www.cv.nrao.edu/~sransom/presto/}}~\citep{2001PhDT.......123R}
pulsar analysis package in order to find and mask RFI found in the data.
The data are read in for each radio frequency channel in 2-second increments. For each of the
2-second segments, the time domain mean and standard deviation are calculated, and a power spectrum
is computed via Fourier transform and the highest Fourier power (with non-zero frequency) is recorded.
For the time domain analysis, any segment for which the mean or standard deviation deviated by more
than 10-sigma from that parameter’s median value over the entire observation is flagged to be masked
during the remaining data analysis. Similarly, if a periodic signal was found in the power spectrum at
a level of greater than 4-sigma, that segment was also marked to be masked.  If more than 70\% of
channels are flagged in a single time interval, then all the channels are masked for that time interval.
Similarly, if more than 30\% of the time intervals are flagged in a single channel then all of that
time intervals for that channel are flagged. For observations processed to date, the
median masking fraction is 1.93\% and less than 2\% of the pointings have required
masking fractions greater than 30\%. In addition to the above RFI excision, known sources of RFI, such as
the 60 Hz AC power line signal, are explicitly removed from the frequency domain after the de-dispersion
step described in the following subsection and after the time series has been Fourier transformed.

\subsection{Dispersion Removal}\label{subsec:dedisp}
After initial RFI masking, the data are de-dispersed at 17,352 trial DMs ranging from
0 to 500 $\mathrm{pc\,cm^{-3}}$. This set of DMs is created such that smearing due to an incorrect DM is less
than the smearing within the frequency channels. Also, when the smearing time within the frequency
channels is $2^N$ times the sample time, the time series is downsampled by a factor of $2^N$. Due to the
discovery of Fast Radio Bursts~\citep{2007Sci...318..777L,2013Sci...341...53T}, which are transients that
may be of extra-galactic origin, we have recently increased our maximum DM. Beginning
in 2014, we increased our DM range from 0 to 500 $\mathrm{pc\,cm^{-3}}$ to 0 to 3000 $\mathrm{pc\,cm^{-3}}$
which now includes 26,532 trial DMs. We are searching all new data with this larger
DM range and plan to re-process all of the already processed data with this new DM range. At 350 MHz,
the dispersive smearing within a frequency channel for DMs of 500, 1000, and 3000 $\mathrm{pc\,cm^{-3}}$
is 2.4, 4.7, and 14.2 ms, respectively.

\subsection{Periodic Search}\label{subsec:search}
Each of the de-dispersed time series is then Fourier transformed and the frequencies and harmonics
associated with known sources of RFI are removed. The resulting time series are then searched for periodicities
using the {\tt accelsearch} tool from {\tt PRESTO}. Two searches are performed, one allowing no
acceleration search and one allowing signals to have a maximum frequency bin drift of
$\pm\frac{50}{n_\mathrm{harm}}$ bins, where $n_\mathrm{harm}$ is the largest harmonic at which the signal was
detected~\citep{2002AJ....124.1788R}. For the low-acceleration search, up to 16 harmonics are
summed and for the high-acceleration search, up to 8 harmonics are summed (in powers-of-two harmonics).
The candidates for each DM and for each of the two acceleration
searches are stored for comparison after all DMs have been searched.

\begin{figure}[h!]
\centering
\includegraphics[width=0.4\textwidth,angle=270]{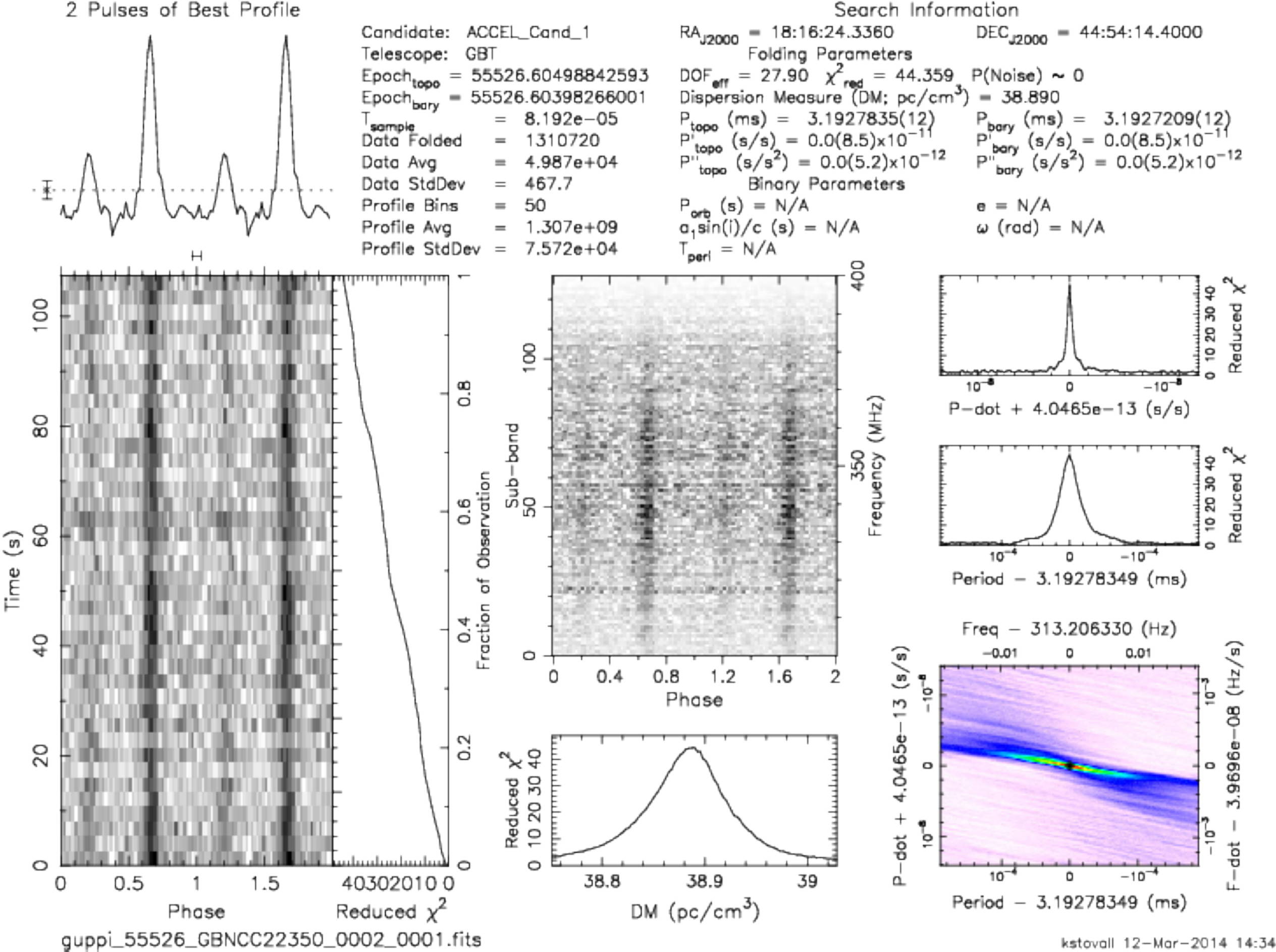}
\caption[Diagnostic Plot of PSR J1816$+$4510]{An example periodic diagnostic plot from the GBNCC pipeline.
This plot shows the GBNCC-discovered PSR J1816$+$4510.}
\label{fig:J1816}
\end{figure}

Once all DM trials have been searched, the periodicity candidates for the two acceleration searches
are gathered into two separate lists. Within each list, candidates within $\pm1.1$ Fourier bins of
each other are combined into
a single candidate and harmonically related candidates are removed. The resulting list
of candidates is passed through a filter which removes candidates which only appeared at
one DM. Then the candidate lists from the two acceleration searches are compared to one
another so that duplicate candidates can be removed (preference is given to the candidate
with higher significance). Candidates whose power in the Fourier domain are greater than
6$\mathrm{\sigma}$ above the Gaussian noise level are folded,
with up to 20 candidates being folded from the search with no acceleration and 10
candidates folded from the high acceleration searches. The candidates are
Folded using the {\tt PRESTO} tool {\tt prepfold}, which creates diagnostic plots for human inspection.
An example of the resulting diagnostic plot is shown in Figure~\ref{fig:J1816}.

\subsection{Single-Pulse Search}\label{subsec:single}

\begin{figure}[h!]
\centering
\includegraphics[width=0.5\textwidth]{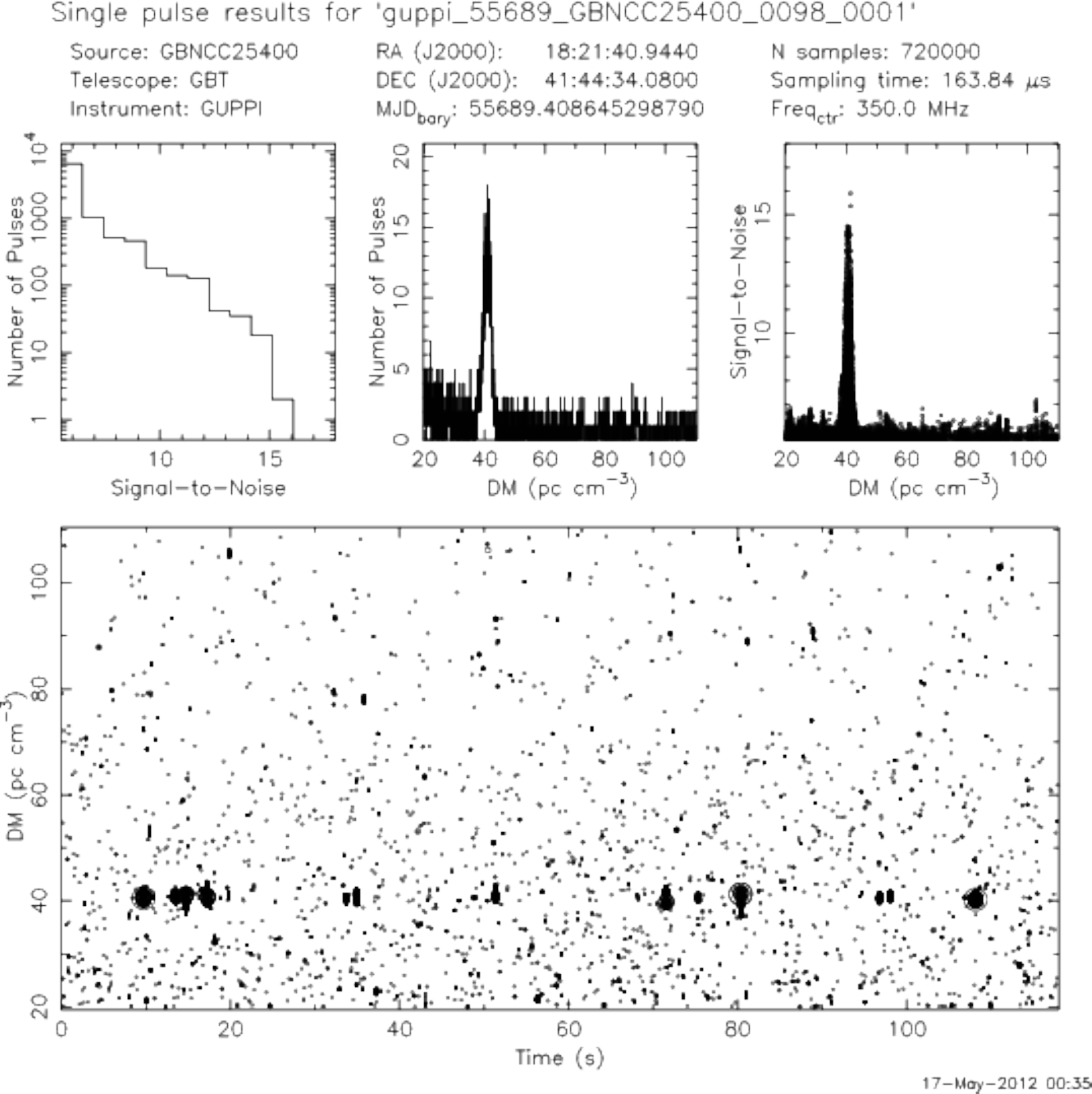}
\caption[Single-Pulse Plot of PSR J1821$+$41]{A single-pulse plot from the GBNCC pipeline. The
signal resulting from pulses of astrophysical origin can be seen at a DM of
40~$\mathrm{pc\,cm^{-3}}$ as this particular beam contains the GBNCC-discovered PSR J1821$+$41. Typical
detections of RRATs look similar, but generally have fewer detected pulses than shown here and in some
cases only a single pulse may be detected. The change in the density of points at around
70~$\mathrm{pc\,cm^{-3}}$ is due to the down-sampling factor changing by a factor of 2.}
\label{fig:single}
\end{figure}

Each of the de-dispersed time series is also searched for single pulses using a
box car matched-filtering algorithm using the {\tt PRESTO} tool {\tt single\_pulse\_search.py}.
The algorithm compares the data to boxcars with widths ranging from 81.92 $\mathrm{\mu s}$
to 100 $\mathrm{ms}$. Pulses with {\bfref SNRs} greater than 5 are stored. {\bfref Though the
lowest SNR RRAT confirmed so far has an SNR of 10, we save pulses down to SNRs of 5 to establish
the noise floor and to provide additional information useful for algorithms used to sift through
potential single pulse candidates. Also, RRATs have been found in previous surveys with SNRs of
about 5, including some of the originally discovered RRATs~\citep[][McLaughlin, private communication]{2006Natur.439..817M}.}
{\bfref After} all de-dispersed time series have been searched, 4 plots are created for human
inspection. Plots are created for the following DM ranges: 0-30,
20-110, 100-310, and 300-500 $\mathrm{pc\,cm^{-3}}$. An example single-pulse plot is shown
in Figure~\ref{fig:single}.

The single-pulse output is then processed through a sifting algorithm (Karako-Argaman et al. 2014, in prep.),
which compares clusters of single-pulse events to the behavior expected of astrophysical sources. The
algorithm then identifies the most promising candidates, thereby significantly reducing the number of
plots to be visually inspected.

\subsection{Candidate Inspection}\label{subsec:candins}
A major obstacle for a large-scale pulsar survey such as the GBNCC is the
inspection of the millions of candidates which are generated. To date, The GBNCC
survey has generated over 1.2 million candidate plots. The GBNCC
team has used a combination of computing algorithms and the help of high school and
undergraduate students to analyze the large volume of candidates. Until recently, the data processed at
each facility has been kept separate and analyzed individually. The candidates resulting
from data processed by the Guillimin cluster have been inspected using a cursory examination based on
signal strength. {\bfref About 200,000 candidates have been generated} from data processed on TACC
resources {\bfref and} were entered into a web based candidate rating tool called ARCC Explorer (Stovall et al., in prep.).
These candidates were pre$-$sorted using the PEACE algorithm~\citep{2013MNRAS.433..688L}
and all {\bfref 200,000} candidates have been examined {\bfref by at least one person with extensive
experience looking at pulsar candidate plots or at least three students, who can have varying experience
looking at such candidate plots. Though it is difficult to determine whether or not all pulsars in this
set have been found, it is unlikely that an undiscovered pulsar with SNR above about 8 exists in the candidates.}
Beginning in late 2013, the results
from each of the processing sites have begun to be combined into one complete set available
to all members of the collaboration hosted by the CyberSKA infrastructure~\footnote{\url{http://www.cyberska.org/}}.
This combined data set is being thoroughly examined by eye and we are employing computer algorithms
such as PEACE and image pattern recognition software described
in~\cite{2014ApJ...781..117Z} to pre-sort the candidates to find new pulsars as quickly as possible.

\section{Re-detections}\label{sec:gbnccredet}
About 80\% of the data analysis for the section of sky north of $\delta=38^\circ$ has been completed. Inspection of
the resulting candidates is ongoing, but to date, we have detected 75 of the known pulsars in this region
of the sky. {\bfref Table~\ref{tab:redet} in the Appendix contains a list of the re-detected sources, their distance
from the center of the GBNCC beam (r), and the observed mean flux density at 350 MHz ($\mathrm{S_{350}^{o}}$) of the
detection}. The ATNF database
lists a total of 158 known pulsars in this region of the sky. However, this number contains many pulsars
which are too weak to be detected in our survey as well as 15 sources that have not been detected at radio
frequencies.

We evaluated whether or not our our theoretically estimated sensitivity is comparable to our
achieved sensitivity using the following method. First, we identified beams in which a known pulsar
was within the 18$'$ HWHM of the 350-MHz beam as taken from The Proposer's Guide for the Green Bank
Telescope~\footnote{\url{https://science.nrao.edu/facilities/gbt/proposing/GBTpg.pdf}}. Then we identified
which of these sources had a mean flux density reported in the ATNF database at either 400 MHz ($\mathrm{S_{400}}$) or
1400 MHz ($\mathrm{S_{1400}}$). For each source with one of these mean flux densities reported, we extracted $\mathrm{S_{400}}$
and $\mathrm{S_{1400}}$. We then estimated a theoretical mean flux density at 350 MHz ($\mathrm{S_{350}^{t}}$) by calculating the
spectral index for the source from $\mathrm{S_{400}}$ and $\mathrm{S_{1400}}$, if both values were available.
We used this spectral index to estimate $\mathrm{S_{350}^{t}}$. In cases with only one mean flux density value available in the
ATNF database, we used a spectral index of -1.7 to scale the reported value to 350 MHz. There can be large uncertainties
in the measurements for $\mathrm{S_{400}}$ and $\mathrm{S_{1400}}$
in the ATNF database, so our resulting $\mathrm{S_{350}^{t}}$ is a very rough estimate of the theoretical mean flux density at
350 MHz. We then took the observed SNR (SNR$_\mathrm{o}$) and used Equation~\ref{eqn:flux} to get an observed
mean flux density ($\mathrm{S_{350}^{o}}$) for each source using the parameters of our survey and $\mathrm{W_{50}}$ for the source
as reported by  the ATNF database. For cases without a reported $\mathrm{W_{50}}$, we fit a set of gaussians to the profile
from our detection and used the FWHM from these gaussians as $\mathrm{W_{50}}$. We also adjusted the value for
$T_\mathrm{sys}$ for each sky location by taking the value for $T_\mathrm{sky}$ at 408 MHz reported in~\cite{1982A&AS...47....1H}
and scaling to 350 MHz using the spectral index -2.6.

\begin{figure}[h!]
\centering
\includegraphics[width=0.5\textwidth]{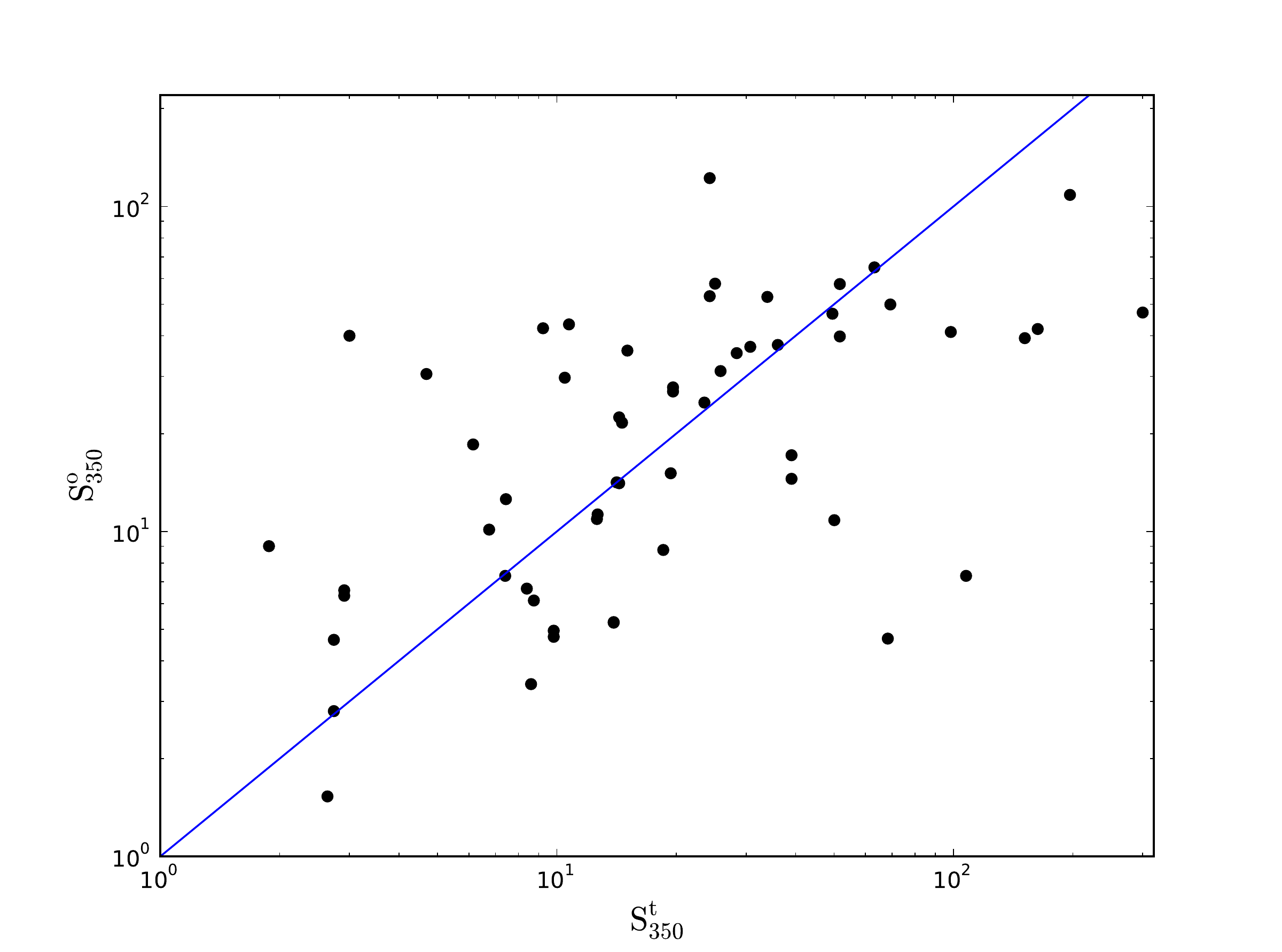}
\caption[Theoretical $\mathrm{S_{350}^{t}}$ versus Observed $\mathrm{S_{350}^{o}}$]{The theoretical mean flux density at 350
MHz {\bfref ($\mathrm{S_{350}^{t}}$)} versus the observed mean flux density at 350
MHz {\bfref ($\mathrm{S_{350}^{o}}$)} for 59 detections of known pulsars with a measured mean flux density at either 1400 MHz
or 400 MHz reported in the ATNF
catalog and that are within 18$'$ of the center of a GBNCC pointing. The blue line represents where
$\mathrm{S_{350}^{t}}$=$\mathrm{S_{350}^{o}}$.}
\label{fig:S350scatter}
\end{figure}

Comparison of $\mathrm{S_{350}^{t}}$ to $\mathrm{S_{350}^{o}}$ revealed that 37\% of the 59 detections satisfying
our requirements were detected with a $\mathrm{S_{350}^{o}}$ larger than expected, another 22\% were within a factor
of 2 of $\mathrm{S_{350}^{t}}$, and a total of 78\% were within a factor of 4. Figure~\ref{fig:S350scatter} shows a
scatter plot of $\mathrm{S_{350}^{t}}$ and $\mathrm{S_{350}^{o}}$. In addition to the rough estimate of the spectral
properties for each pulsar, there are many additional factors which can affect individual
observations including but not limited to interstellar scintillation, pulsar nulling, and RFI. Another
factor, in the case of slow pulsars, is that our integration time of 120-s does not allow for a large number
of pulses, so normal pulse-to-pulse variability can affect the observed SNR. Of the 59 detections,
there were 6 observations for which the ratio of $\mathrm{S_{350}^{t}}$/$\mathrm{S_{350}^{o}}$ was 8 or larger. Visual
inspection of these six detections revealed that three of them showed evidence of nulling and the other three had
intensities which were variable throughout the observation. Also, three of them were near the 18$'$ HWHM of the 350 MHz
receiver and one was severely affected by RFI. We conclude that our expected to true sensitivity are roughly in agreement,
given all of the assumptions involved.

\section{Pulsar Timing Analysis}~\label{sec:timing}
In Section~\ref{sec:gbnccdisc}, we present timing solutions for five new GBNCC discoveries. Observations
were conducted using a combination of GBNCC survey time at 350 MHz, dedicated follow-up projects 
(GBT11B-070 and GBT13A-458) at 820 MHz, and observations from a few smaller projects and
in one case from NANOGrav observations. The observations taken during GBNCC survey time were
typically 10 to 15 minutes in duration and use the same
GBNCC observing setup described in Section~\ref{sec:gbnccdesc}. The 820 MHz observations of
PSR J1816$+$4510 under project GBT11B-070 were typically 15 to 30 minutes in duration, while the
observations of the pulsars in GBT13A-458 were typically 4-5 minutes in duration.
The observations under GBT11B-070 and GBT13A-458 at 820 MHz had a bandwidth of 200 MHz divided into 2,048 frequency channels and
a sample time of 40.96 $\mu s$. Additional TOAs for PSR J0645$+$5158 were obtained from the
NANOGrav collaboration due to it being a promising new candidate for inclusion in PTAs.
The NANOGrav data consist of two sets of data, one at a center frequency of 820 MHz with
200 MHz of bandwidth divided into 128 coherently de-dispersed frequency bins, and one at a center
frequency of 1500 MHz with 800 MHz of bandwidth divided into 512 coherently de-dispersed frequency
bins. Both data sets were taken in fold mode with 2,048 bins across the pulse profile and profiles
were written out every 15 seconds. Two complete orbits (one at 1500 MHz and the other at 820 MHz) of
PSR J1816$+$4510 were observed under the Fermi project GLST051395. This data was taken in the
same mode as the NANOGrav data set described above. We also observed PSR J0645$+$5158 under the
project GBT11A-075 and PSR J0636$+$5129 under project GBT12A-473 to reduce the positional uncertainty
of each source. These observations consisted of 820 MHz, 1500 MHz, and 2200 MHz observations. The
820 MHz observations were taken in the same mode as our main timing observations, while the higher
frequency observations were each taken with 800 MHz of bandwidth across 2,048 frequency channels
with a sample time of 40.96 $\mu s$.

For each of the pulsars, we created template profiles by first summing together multiple observations and
then fitting the resulting high signal-to-noise ratio profiles to Gaussians using a least squares minimization
algorithm\footnote{{\tt PRESTO}'s {\tt pygaussfit.py}}. For each pulsar, the number of Gaussian
components needed to represent the profile did not change from one frequency to another, however
the FWHM did change, so we used the 820 MHz profile, but modified the FWHM for other frequencies. Using
the template profile, we calculated
times of arrival (TOAs) of pulses using a least squares fit in the Fourier
domain~\citep{1992PTRSL.341..117T}\footnote{{\tt PRESTO}'s {\tt get\_TOAs.py} which uses the {\tt FFTFIT} routine}.
In the case of the isolated pulsars and the
long-period binary pulsar J0214$+$5222, we divided the observations into two frequency bands and
obtained two TOAs from each band for a total of four TOAs per observation. For the short-period binary
pulsars, we initially summed the frequency channels into a single sub-band and obtained 5 TOAs from each
observation in order to sample more of the pulsar's orbital phase. Once we solved these pulsars' orbits,
we then divided the observation into two frequency sub-bands and obtained two TOAs from each sub-band. We
used these TOAs to obtain a timing solution by accounting for every rotation of the pulsar using the
{\tt TEMPO}{\bfref \footnote{\url{http://tempo.sourceforge.net}}} software package.

Each of our timing solutions use the DE421 Solar System ephemeris and TT(BIPM) clock corrections.
For each pulsar, we obtained good timing solutions with no systematic trends. For four of the
five pulsars, we multiplied each TOA uncertainty by an error factor (values range from 1.0 to 1.5)
in order to get a reduced $\chi^2$ near one. This ensures that the fitted parameters and their errors
are justified by the actual TOA rms. We did not do this for PSR J0214$+$5222 because several parameters
are still highly covariant due to sparse coverage of the orbit, so its $\chi^2$ is less than one.

\section{Discoveries}\label{sec:gbnccdisc}
Prior to the ongoing full data analysis, we performed a preliminary search of the GBNCC
data by selecting beams containing a gamma-ray point source as identified by \textit{Fermi}.
This method led to the search of 128 of the $\approx30,000$
beams from stage 1 of the GBNCC survey. This search resulted in the
detection of what appeared to be two new MSPs. Comparison with recent \textit{Fermi} Pulsar Search
Consortium discoveries revealed that one of them (J2302$+$4442) had been previously
discovered~\citep{2011ApJ...732...47C}, but the other (J1816$+$4510) was a previously unknown MSP. PSR
J1816$+$4510 is described in detail below.

After this preliminary data analysis was complete, we began processing the data in earnest,
which as of 2014 March 18 has resulted in the discovery of 67 pulsars, including 9 MSPs and 6
intermediate-spin-period pulsars. Additionally, 7 RRATs have been discovered which will be presented
in Karako-Argaman et al. 2014, in prep. Figure~\ref{fig:gbncc_skycoverage} shows
our discovered sources on the sky. We highlight that we are finding MSPs spread across the
sky, which is essential for the efforts of the PTAs to detect the stochastic background from GW
sources.

\setcounter{table}{1}
\begin{table*}[h]
\caption{ Discovery parameters for 62 GBNCC discovered pulsars. \label{tab:gbnccdisc} }
\begin{center} {\footnotesize
\begin{tabular}{cccccccc}
Pulsar & P & DM & RA & Dec & $D$\tablenotemark{a} & $\mathrm{SNR_{Disc}}$ & $\mathrm{S_{350}^{o}}$ \\
   & s & $\mathrm{pc\, cm^{-3}}$ & hh:mm & dd:mm &  kpc & & mJy \\
\hline
J0034$+$69 & 0.03680412(16) & 80.0(3) & 00:34(3) & 69:43(18) & 2.8 & 61.2 & 4.5 \\
J0053$+$69 & 0.832938(28) & 117(2) & 00:53(3) & 69:39(18) & 4.3 & 676.9 & 30.5 \\ 
J0059$+$50 & 0.996009(61) & 67(2) & 00:59(2) & 50:02(18) & 2.7 & 120.0 & 5.6 \\ 
J0112$+$66 & 4.30124(70) & 112(3) & 01:12(3) & 66:22(18) & 3.4 & 340.1 & 13.6 \\ 
J0125$+$62 & 1.708233(98) & 118(2) & 01:26(3) & 62:35(18) & 3.9 & 50.9 & 3.4 \\ 
\noalign{\smallskip}
J0136$+$63 & 0.717895(27) & 286(2) & 01:36(3) & 63:42(18) & $>$44.3\tablenotemark{b} & 40.3 & 3.3 \\ 
J0141$+$63 & 0.04668588(49) & 272.6(4) & 01:41(2) & 63:08(8) & $>$44.3\tablenotemark{b} & 13.1 & 1.3 \\ 
J0213$+$52 & 0.376384(15) & 38(2) & 02:13(2) & 52:32(8) & 1.5 & 67.6 & 4.1 \\ 
J0325$+$67 & 1.364741(62) & 65(2) & 03:26(3) & 67:49(18) & 2.3 & 816.4 & 26.3 \\ 
J0338$+$66 & 1.76200(15) & 67(2) & 03:39(3) & 66:44(18) & 2.2 & 260.2 & 10.5 \\ 
\noalign{\smallskip}
J0358$+$42 & 0.2264777(50) & 46(1) & 03:58(1) & 42:06(9) & 1.6 & 84.0 & 5.5 \\ 
J0358$+$66 & 0.09150682(63) & 62.3(4) & 03:58(3) & 66:40(18) & 2.2 & 103.8 & 5.3 \\ 
J0417$+$61 & 0.440283(13) & 71(2) & 04:17(3) & 61:08(18) & 2.3 & 31.4 & 1.5 \\ 
J0510$+$38 & 0.07656440(44) & 69(1) & 05:09:59(40) & 38:12(8) & 1.9 & 33.1 & 3.4 \\ 
J0519$+$54 & 0.340194(12) & 43(1) & 05:20(2) & 54:25(18) & 1.5 & 63.7 & 3.8 \\ 
\noalign{\smallskip}
J0610$+$37 & 0.443861(14) & 39(2) & 06:11(2) & 37:18(18) & 1.2 & 29.0 & 1.3 \\ 
J0614$+$83 & 1.039203(51) & 44(1) & 06:14(10) & 83:14(18) & 2.2 & 57.8 & 1.9 \\ 
J0645$+$80 & 0.657873(24) & 50(3) & 06:46(7) & 80:09(18) & 2.8 & 119.0 & 5.5 \\ 
J0737$+$69 & 6.82424(63) & 16(3) & 07:37(2) & 69:14(8) & 0.73 & 59.4 & 1.2 \\ 
J0740$+$41 & 0.0031392150(48) & 20.83(9) & 07:41(1) & 41:04(8) & 0.71 & 47.6 & 5.8 \\ 
\noalign{\smallskip}
J0741$+$66 & 0.00288570374(45) & 14.96(4) & 07:42(2) & 66:20(8) & 0.67 & 365.1 & 32.5 \\ 
J0746$+$66 & 0.4076702(93) & 28(1) & 07:46(3) & 66:36(18) & 1.4 & 273.3 & 7.4 \\ 
J0750$+$57 & 1.174875(65) & 27(2) & 07:50(2) & 57:00(18) & 1.0 & 55.8 & 2.9 \\ 
J0943$+$41 & 2.229489(99) & 21(2) & 09:43(2) & 41:09(18) & 0.79 & 550.1 & 8.6 \\ 
J1101$+$65 & 3.63132(53) & 19(3) & 11:02(3) & 65:07(18) & 0.82 & 509.2 & 10.4 \\ 
\noalign{\smallskip}
J1110$+$58 & 0.793348(21) & 26(1) & 11:11(3) & 58:52(18) & 1.3 & 65.9 & 1.6 \\ 
J1122$+$78 & 0.0042013838(14) & 11.22(8) & 11:25:58(9) & 78:23(2) & 0.62 & 274.5 & 17.1 \\ 
J1320$+$67 & 1.028620(35) & 28(2) & 13:20(3) & 67:30(18) & 1.5 & 33.7 & 1.4 \\ 
J1627$+$86 & 0.3957850(87) & 46(1) & 16:27(5) & 86:54(4) & 3.0 & 199.9 & 5.6 \\ 
J1629$+$43 & 0.1811729(16) & 7.3(8) & 16:29(2) & 43:59(18) & 0.61 & 99.4 & 3.2 \\ 
\noalign{\smallskip}
J1647$+$66 & 1.599819(50) & 23(2) & 16:48(1) & 66:04(8) & 1.2 & 147.1 & 3.3 \\ 
J1649$+$80 & 0.0020210978(13) & 31.09(4) & 16:50(4) & 80:45(8) & 1.6 & 69.7 & 4.1 \\ 
J1706$+$59 & 1.476687(58) & 31(2) & 17:07(3) & 59:10(18) & 1.8 & 369.6 & 8.0 \\ 
J1710$+$49 & 0.0032202231(21) & 7.09(8) & 17:10:29(13) & 49:20(2) & 0.66 & 117.4 & 14.1 \\ 
J1800$+$50 & 0.578364(15) & 23(2) & 18:01(2) & 50:28(18) & 1.4 & 702.6 & 23.5 \\ 
\noalign{\smallskip}
J1815$+$55 & 0.426802(13) & 59(2) & 18:15(2) & 55:29(18) & $>$50.0\tablenotemark{b} & 108.2 & 4.2 \\ 
J1821$+$41 & 1.261787(46) & 40(3) & 18:22(3) & 41:45(18) & 2.5 & 115.5 & 2.6 \\ 
J1859$+$76 & 1.393617(87) & 47(2) & 18:59(6) & 76:54(18) & 2.9 & 106.3 & 3.3 \\ 
J1911$+$37 & 0.851095(33) & 72(2) & 19:11(2) & 37:38(18) & 4.2 & 216.6 & 7.0 \\ 
J1921$+$42 & 0.595201(15) & 53(1) & 19:21:56(40) & 42:25(8) & 3.2 & 180.9 & 8.1 \\ 
\noalign{\smallskip}
J1922$+$58 & 0.529623(10) & 53(1) & 19:22(3) & 58:28(18) & 3.3 & 79.0 & 3.1 \\ 
J1929$+$62 & 1.456004(66) & 67.7(7) & 19:29(3) & 62:16(18) & 6.4 & 38.1 & 1.4 \\ 
J1935$+$52 & 0.568387(17) & 71(2) & 19:35(2) & 52:12(18) & 4.4 & 52.3 & 3.1 \\ 
J1939$+$66 & 0.022260606(52) & 41.2(1) & 19:40(3) & 66:12(18) & 2.3 & 46.4 & 2.1 \\ 
J1941$+$43 & 0.840887(35) & 79(2) & 19:42(2) & 43:23(18) & 4.3 & 213.6 & 9.1 \\ 
\noalign{\smallskip}
J1942$+$81 & 0.2035682(44) & 40.3(5) & 19:42(3) & 81:06(8) & 2.1 & 175.9 & 4.8 \\ 
J1953$+$67 & 0.008565431(18) & 57.2(1) & 19:53(3) & 67:02(18) & 3.4 & 29.5 & 1.7 \\ 
J1954$+$43 & 1.386961(94) & 130(3) & 19:55(2) & 43:50(18) & 6.9 & 22.5 & 2.8 \\ 
J2001$+$42 & 0.719161(22) & 55(1) & 20:02(3) & 42:43(18) & 3.3 & 321.0 & 20.2 \\ 
J2017$+$59 & 0.403619(12) & 61(3) & 20:18(3) & 59:13(18) & 3.2 & 10.5 & 1.1 \\ 
\noalign{\smallskip}
J2027$+$74 & 0.515229(14) & 11(3) & 20:28(5) & 74:47(18) & 0.9 & 30.7 & 2.4 \\ 
J2105$+$28 & 0.405737(11) & 62.4(8) & 21:06(2) & 28:29(18) & 3.7 & 249.8 & 8.6 \\ 
J2113$+$67 & 0.5521697(88) & 55.1(8) & 21:14(3) & 67:02(18) & 2.7 & 26.6 & 1.0 \\ 
J2122$+$54 & 0.1388657(13) & 31.7(7) & 21:23(1) & 54:33(8) & 2.1 & 40.1 & 3.4 \\ 
J2137$+$64 & 1.75087(14) & 106(2) & 21:37(3) & 64:19(18) & 4.6 & 167.2 & 6.4 \\ 
\noalign{\smallskip}
J2205$+$62 & 0.3227871(63) & 167(3) & 22:06(3) & 62:04(18) & 6.7 & 8.6 & 1.2 \\ 
J2207$+$40 & 0.6369852(85) & 12(2) & 22:07(2) & 40:57(18) & 1.0 & 59.5 & 3.8 \\ 
J2210$+$57 & 2.05743(13) & 189.43(6) & 22:11(3) & 57:29(18) & 6.1 & 9.9 & 1.1 \\
J2229$+$64 & 1.89312(13) & 194(1) & 22:29(3) & 64:58(18) & $>$46.9\tablenotemark{b} & 56.1 & 3.4 \\ 
J2243$+$69 & 0.855405(32) & 68(2) & 22:44(7) & 69:40(18) & 2.9 & 78.8 & 3.2 \\ 
\noalign{\smallskip}
J2316$+$69 & 0.813386(22) & 71(2) & 23:17(3) & 69:12(18) & 2.8 & 51.9 & 2.5 \\ 
J2353$+$85 & 1.011691(37) & 38(2) & 23:54(7) & 85:34(8) & 1.9 & 183.6 & 4.9 \\ 
\hline
\end{tabular} }
\tablenotetext{1}{Distances are from the NE2001 electron model with $\sim20\%$ uncertainty in distance values.~\citep{2002astro.ph..7156C}}
\tablenotetext{2}{DM values are slightly larger than the expected highest DM given by the NE2001 electron model and therefore the model does not give reliable distances for these sources. The maximum DM values expected along these lines-of-sight are 212, 205, 51, and 187 $\mathrm{pc\,cm^{-3}}$ for PSRs J0136$+$63, J0141$+$63, J1815$+$55, and J2229$+$64, respectively.}
\end{center}
\end{table*}

\begin{figure}
\centering
\includegraphics[width=0.5\textwidth]{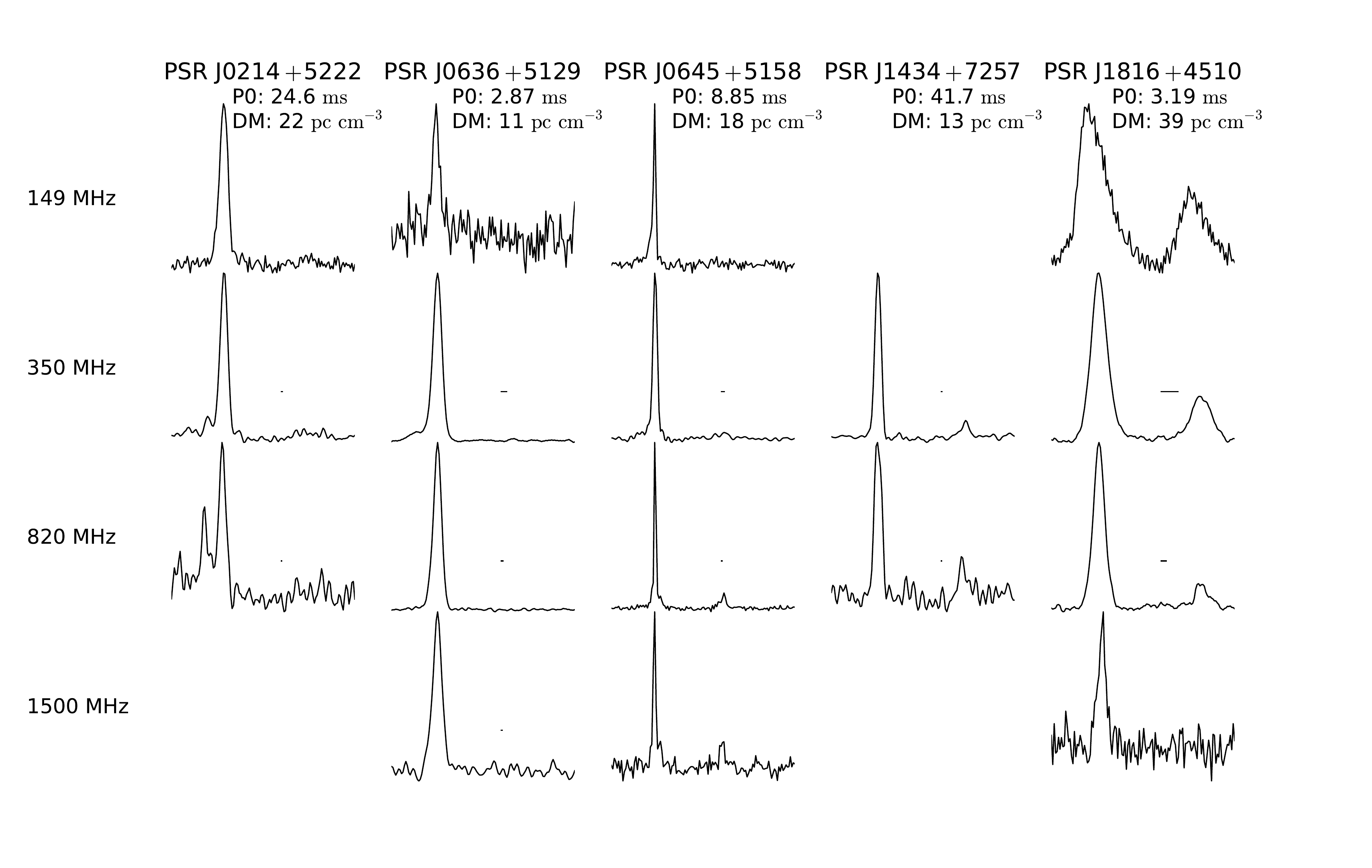}
\caption[Pulse Profiles for GBNCC Pulsars]{Pulse profiles for PSRs J0214$+$5222, J0636$+$5129, J0645$+$5158, J1434$+$7257,
and J1816$+$4510 are shown. Each profile was calculated by summing profiles from multiple observations ({\bfref ranging from 10 minutes
to 1 hour}) at the same frequency and each contains 128 phase bins. {\bfref The 149-MHz profiles were made from LOFAR observations, while
350-MHz, 820-MHz, and 1500-MHz profiles were made from observations using the GBT's GUPPI backend.} The 1500-MHz profile of PSR
J0636$+$5129 and the profiles of all pulsars at 350 MHz and 820 MHz were created with incoherently de-dispersed data, but the
1500-MHz profiles of PSRs J0645$+$5158 and J1816$+$4510 {\bfref as well as all 149-MHz profiles} were created using coherently
de-dispersed data. {\bfref Profiles made from incoherently de-dispersed data have lines showing dispersive smearing within
the frequency channels for that particular profile.}}
\label{fig:gbnccprofile}
\end{figure}

Discovery parameters for 62 discovered pulsars are given in Table~\ref{tab:gbnccdisc}. {\bfref The parameters
included are the observed spin period, DM, position, an estimated distance using the DM, the discovery SNR
($\mathrm{SNR_{Disc}}$), and an estimated mean flux density at 350 MHz ($\mathrm{S_{350}^{o}}$). The mean flux
density was estimated by putting the parameters for the
GBNCC survey, the discovery SNR, and the FWHM from the discovery pulse profile into Equation~\ref{eqn:flux}.}
We present timing solutions for 5 of the pulsars in Tables~\ref{tab:isolated} through \ref{tab:binaries}
and show their pulse profiles at{\bfref 149,} 350 MHz, 820 MHz, and 1500 MHz (where applicable) in Figure~\ref{fig:gbnccprofile}.
{\bfref The profiles at 350 MHz, 820 MHz, and 1500 MHz were made using data described in Section~\ref{sec:timing}, 
while the 149-MHz profiles are from LOFAR. The LOFAR profiles are averages over 10 - 60 minutes, using 78 MHz of bandwidth
centered at 149 MHz.  The data were coherently de-dispersed within each of 400 195-kHz frequency subbands (see~\citet{2011A&A...530A..80S}
for further details of this observing mode).}
Follow-up observations and analysis of the 62 pulsars for which we did not present timing
solutions as well as the 7 RRATs are ongoing and timing solutions will appear in future publications.

Two of the pulsars (J0645$+$5158 and J1434$+$7257) are isolated and their timing solutions are
presented in Table~\ref{tab:isolated}. The other three pulsars
(J0214$+$5222, J0636$+$5129, and J1816$+$4510) are binary pulsars. In the case of J0636$+$5129 and
J1816$+$4510, the orbits are short-period and highly circular, so we have used the ELL1 {\tt tempo}
model~(see Appendix of \citealt{2001MNRAS.326..274L}).
PSR J0214$+$5222 is in a very long-period orbit, which is more eccentric and therefore we have used
the BT {\tt tempo} model~\citep{1976ApJ...205..580B}. The solution for the long-period binary, J0214$+$5222, is given in
Table~\ref{tab:J0214} and the solutions for the short-period binaries are given in Table~\ref{tab:binaries}.
{\bfref Figure~\ref{fig:gbncc_massvpb}, described in Section~\ref{sec:gbnccintro}, shows the minimum mass and
orbital periods of our short-period binary discoveries plotted among previously known binary pulsars with
spin periods below 10 ms.} This plot
shows how our discovered binaries (blue stars) compare to the known populations of MSPs with white
dwarf companions (unfilled, black circles) as well as redback (red circles), black widow systems
(black circles), and the ``diamond planet'' systems (black diamonds). Additional details
about our discovered systems are given in the subsections below.

\begin{table*}[h]
  \begin{center} {\footnotesize
\caption[Timing solutions for 2 GBNCC discovered isolated pulsars.]{Timing solutions and derived parameters for the isolated PSRs J0645$+$5158 and J1434$+$7257.}
\begin{tabularx}{\textwidth}{Xcc}
  Parameter &
   PSR J0645+5158 &
   PSR J1434+7257 \\
    \hline
    \multicolumn{3}{c}{Timing Parameters} \\
    \hline\hline
    Right Ascension (J2000)                       \dotfill & 06:45:59.08191(1)           & 14:33:59.7370(9)            \\
    Proper Motion in RA ($\mathrm{mas\;yr^{-1}}$) \dotfill & 1.2(1)                      & \nodata                       \\
    Declination (J2000)                           \dotfill & 51:58:14.9210(1)            & 72:57:26.512(6)             \\
    Proper Motion in Dec ($\mathrm{mas\;yr^{-1}}$)\dotfill & -7.5(2)                     & \nodata                       \\
    Parallax (mas)                                \dotfill & 1.4(4)                        & \nodata                     \\
    Pulsar Period (s)                          \dotfill & 0.00885349668613151(7)      & 0.04174114783993(3)         \\
    Period Derivative (s s$^{-1}$)                 \dotfill & 4.923(8)$\times$10$^{-21}$  & 5.50(1)$\times$10$^{-19}$   \\
    Dispersion Measure (\dmu)                     \dotfill & 18.247536(9)                & 12.605(1)                   \\
    Reference Epoch (MJD)                         \dotfill & 56143.0                       & 55891.0                       \\
    Span of Timing Data (MJD)                     \dotfill & 55700--56586                  & 55196--56585                  \\
    Number of TOAs                                \dotfill & 3747                          & 127                           \\
    RMS Residual ($\us$)                          \dotfill & 0.51                          & 58                         \\
    350 MHz Flux (mJy)                            \dotfill & 2.4                           & 1.3                         \\
    820 MHz Flux (mJy)                            \dotfill & 0.37                          & 0.34                         \\
    FWHM 350 MHz (ms)                             \dotfill & 0.23                          & 1.6                           \\
    FWHM 820 MHz (ms)                             \dotfill & 0.086                         & 1.9                           \\
    \hline
    \multicolumn{3}{c}{Derived Parameters} \\
    \hline\hline
    Galactic Longitude                            \dotfill & 163.96                        & 113.08                        \\
    Galactic Latitude                             \dotfill & 20.25                         & 42.15                         \\
    Lutz-Kelker Adjusted Parallax (mas)           \dotfill & 1.0(4)                      & \nodata                     \\
    DM-Derived Distance (kpc)                     \dotfill & 0.7                           & 0.7                           \\
    PX-Derived Distance (kpc)                     \dotfill & 0.7$\mathrm{^{+0.2}_{-0.1}}$ & \nodata                       \\
    Shklovskii Adjusted Period Derivative($s\, \ps$) \dotfill & 4.1$\mathrm{^{+0.2}_{-0.3}\times10^{-21}}$ & \nodata         \\
    Surface Magnetic Field Strength ($10^{8}$ Gauss) \dotfill & 1.9  & 48                      \\
    Spindown Luminosity ($10^{32}$ erg/s)         \dotfill & 2.4        & 3.0                          \\
    Characteristic Age (Gyr)                      \dotfill & 34          & 1.2                        \\
\end{tabularx} }
\tablecomments{Numbers in parentheses represent 1-$\sigma$ uncertainties in the
last digits as determined by \texttt{TEMPO}, scaled such that the reduced
$\chi^2 = 1$.  All timing solutions use the DE421 Solar System Ephemeris and the
UTC(BIPM) time system. Derived quantities assume an $R = 10\; \km$ neutron star with
$I = 10^{45}\; \gm\, \cm^2$. The \DM\ derived distances were calculated using the NE2001
model of Galactic free electron density, and have typical errors of $\approx20\%$
\citep{2002astro.ph..7156C}. The parallax and parallax distance was determined using \texttt{TEMPO}
and then corrected for Lutz-Kelker bias using the technique described
in~\cite{2010MNRAS.405..564V}. The Shklovskii adjusted period derivative is corrected for the Shklovskii effect~\citep{1970SvA....13..562S}
using the proper motion and PX-derived distance. The surface magnetic field strength, spindown luminosity, and characteristic
age were calculated using the Shlovskii adjusted period derivative.}
\end{center}
\label{tab:isolated}
\end{table*}

\begin{table}[h]
  \begin{center} {\footnotesize
  \caption[Timing solutions for PSR J0214$+$5222.]{Timing solution and derived parameters for the long-period GBNCC discovered pulsar J0214$+$5222.}
  \begin{tabularx}{0.45\textwidth}{Xc}
  Parameter &
   PSR J0214+5222 \\
    \hline
    \multicolumn{2}{c}{Timing Parameters} \\
    \hline\hline
    Right Ascension (J2000)                       \dotfill & 02:14:55.2713(34)             \\
    Declination (J2000)                           \dotfill & 52:22:40.95(19)               \\
    Pulsar Period (s)                             \dotfill & 0.02457529479071(22)          \\
    Period Derivative (s s$^{-1}$)                 \dotfill & 2.99(10)$\times$10$^{-19}$    \\
    Dispersion Measure (\dmu)                     \dotfill & 22.0354(34)                   \\
    Reference Epoch (MJD)                         \dotfill & 55974.0                       \\
    Span of Timing Data (MJD)                     \dotfill & 55353--56594                  \\
    Number of TOAs                                \dotfill & 69                            \\
    RMS Residual ($\us$)                          \dotfill & 75                           \\
    350 MHz Flux (mJy)                            \dotfill & 0.90                          \\
    820 MHz Flux (mJy)                            \dotfill & 0.24                          \\
    FWHM 350 MHz (ms)                             \dotfill & 1.2                           \\
    FWHM 820 MHz (ms)                             \dotfill & 2.5                           \\
    \hline
    \multicolumn{2}{c}{Binary Parameters} \\
    \hline\hline
    Orbital Period (days)                         \dotfill & 512.0397(3)                 \\
    Projected Semi-major Axis (lt-s)              \dotfill & 174.5658(2)                 \\
    Epoch of Periastron (MJD)                     \dotfill & 56126.61(1)                 \\
    Orbital Eccentricity                          \dotfill & 0.0053283(5)                \\
    Longitude of Periastron (deg)                 \dotfill & 210.59(1)                  \\
    Mass Function (\Msun)                         \dotfill & 0.02178                       \\
    Minimum Companion Mass (\Msun)                \dotfill & 0.4157                        \\
    \hline
    \multicolumn{2}{c}{Derived Parameters} \\
    \hline\hline
    Galactic Longitude                            \dotfill & 135.63                        \\
    Galactic Latitude                             \dotfill & -8.42                         \\
    DM-Derived Distance (kpc)                     \dotfill & 1.0                           \\
    Surface Magnetic Field Strength ($10^{8}$ Gauss) \dotfill & 27                         \\
    Spindown Luminosity ($10^{32}$ erg/s)         \dotfill & 8.0                          \\
    Characteristic Age (Gyr)                      \dotfill & 1.3                        \\
  \tablecomments{The notes for Table~\ref{tab:isolated} also apply to this table.
Minimum companion masses were calculated assuming a $1.4\;\Msun$ pulsar.}
\end{tabularx} }
\end{center}
\label{tab:J0214}
\end{table}

\begin{table*}[h]
  \begin{center} {\footnotesize
  \caption[Timing solutions for 2 GBNCC discovered binary pulsars.]{Timing solutions and derived parameters for the GBNCC discovered binary PSRs J0636$+$5129 and J1816$+$4510.}
  \begin{tabularx}{\textwidth}{Xcc}
  Parameter &
   PSR J0636$+$5129 &
   PSR J1816$+$4510 \\
    \hline
    \multicolumn{3}{c}{Timing Parameters} \\
    \hline\hline
    Right Ascension (J2000)                       \dotfill & 06:36:04.84645(2)             & 18:16:35.93436(7)           \\
    Proper Motion in RA ($\mathrm{mas\;yr^{-1}}$) \dotfill & 4.3(9)                        & 5.3(8)                      \\
    Declination (J2000)                           \dotfill & 51:28:59.9625(6)              & 45:10:33.8618(8)            \\
    Proper Motion in DEC ($\mathrm{mas\;yr^{-1}}$)\dotfill & 2(1)                          & -3(1)                     \\
    Parallax (mas)                                \dotfill & 4.9(6)                        & \nodata                       \\
    Pulsar Period (s)                          \dotfill & 0.0028689528463143(1)         & 0.0031931035538505(2)       \\
    Period Derivative (s s$^{-1}$)                 \dotfill & 3.38(3)$\times$10$^{-21}$     & 4.310(1)$\times$10$^{-20}$  \\
    Dispersion Measure (\dmu)                     \dotfill & 11.10598(6)                   & 38.8874(4)                  \\
    Reference Epoch (MJD)                         \dotfill & 56307.0                       & 56047.0                       \\
    Span of Timing Data (MJD)                     \dotfill & 56028--56586                  & 55508--56586                  \\
    Number of TOAs                                \dotfill & 1180                          & 457                           \\
    RMS Residual ($\us$)                          \dotfill & 2.6                           & 8.7                           \\
    350 MHz Flux (mJy)                            \dotfill & 1.8                           & 1.5                         \\
    820 MHz Flux (mJy)                            \dotfill & 0.66                          & 0.28                         \\
    FWHM 350 MHz (ms)                             \dotfill & 0.16                          & 0.29                          \\
    FWHM 820 MHz (ms)                             \dotfill & 0.15                          & 0.24                         \\
    \hline
    \multicolumn{3}{c}{Binary Parameters} \\
    \hline\hline
    Orbital Period (days)                         \dotfill & 0.0665513390(1)               & 0.3608934817(2)             \\
    Projected Semi-major Axis (lt-s)              \dotfill & 0.0089859(1)                  & 0.595405(1)                 \\
    Epoch of Ascending Node (MJD)                 \dotfill & 56307.4294776(5)              & 56047.5490643(2)           \\
    Eccentricity Sin(Omega)                       \dotfill & -1(3)$\times$10$^{-5}$        & 5(3)$\times$10$^{-6}$     \\
    Eccentricity Cos(Omega)                       \dotfill & -2(3)$\times$10$^{-5}$        & 6(4)$\times$10$^{-6}$     \\
    Mass Function (\Msun)                         \dotfill & 1.759$\times$10$^{-7}$        & 0.00174                       \\
    Minimum Companion Mass (\Msun)                \dotfill & 0.007035                      & 0.1619                        \\
    \hline
    \multicolumn{3}{c}{Derived Parameters} \\
    \hline\hline
    Galactic Longitude                            \dotfill & 163.91                        & 72.83                         \\
    Galactic Latitude                             \dotfill & 18.64                         & 24.74                         \\
    Lutz-Kelker Adjusted Parallax (mas)           \dotfill & 4.51$\mathrm{^{+0.57}_{-0.58}}$ & \nodata                     \\
    DM-Derived Distance (kpc)                     \dotfill & 0.5                           & 2.4                           \\
    PX-Derived Distance (kpc)                     \dotfill & 0.203$\mathrm{^{+0.027}_{-0.021}}$ & \nodata                       \\
    Shklovskii Adjusted Period Derivative($s\, \ps$) \dotfill & 3.373$\mathrm{^{+0.044}_{-0.057}\times10^{-21}}$ & 4.180$\mathrm{^{+0.077}_{-0.121}\times10^{-20}}$         \\
    Surface Magnetic Field Strength ($10^{8}$ Gauss) \dotfill & 1.0                        & 4.0                     \\
    Spindown Luminosity ($10^{32}$ erg/s)         \dotfill & 56                          & 510                         \\
    Characteristic Age (Gyr)                      \dotfill & 14                          & 1.2                        \\
  \tablecomments{The notes for Tables~\ref{tab:isolated} and ~\ref{tab:J0214} also apply to this table. The Shklovskii adjusted
  period derivative for PSR J1816$+$4510 was calculated using the distance from~\cite{2013ApJ...765..158K} since we do not have a
  measurable parallax.}
\end{tabularx} }
\end{center}
\label{tab:binaries}
\end{table*}

\subsection{PSR J0214$+$5222}\label{subsec:J0214}
\begin{figure}
\includegraphics[width=0.5\textwidth]{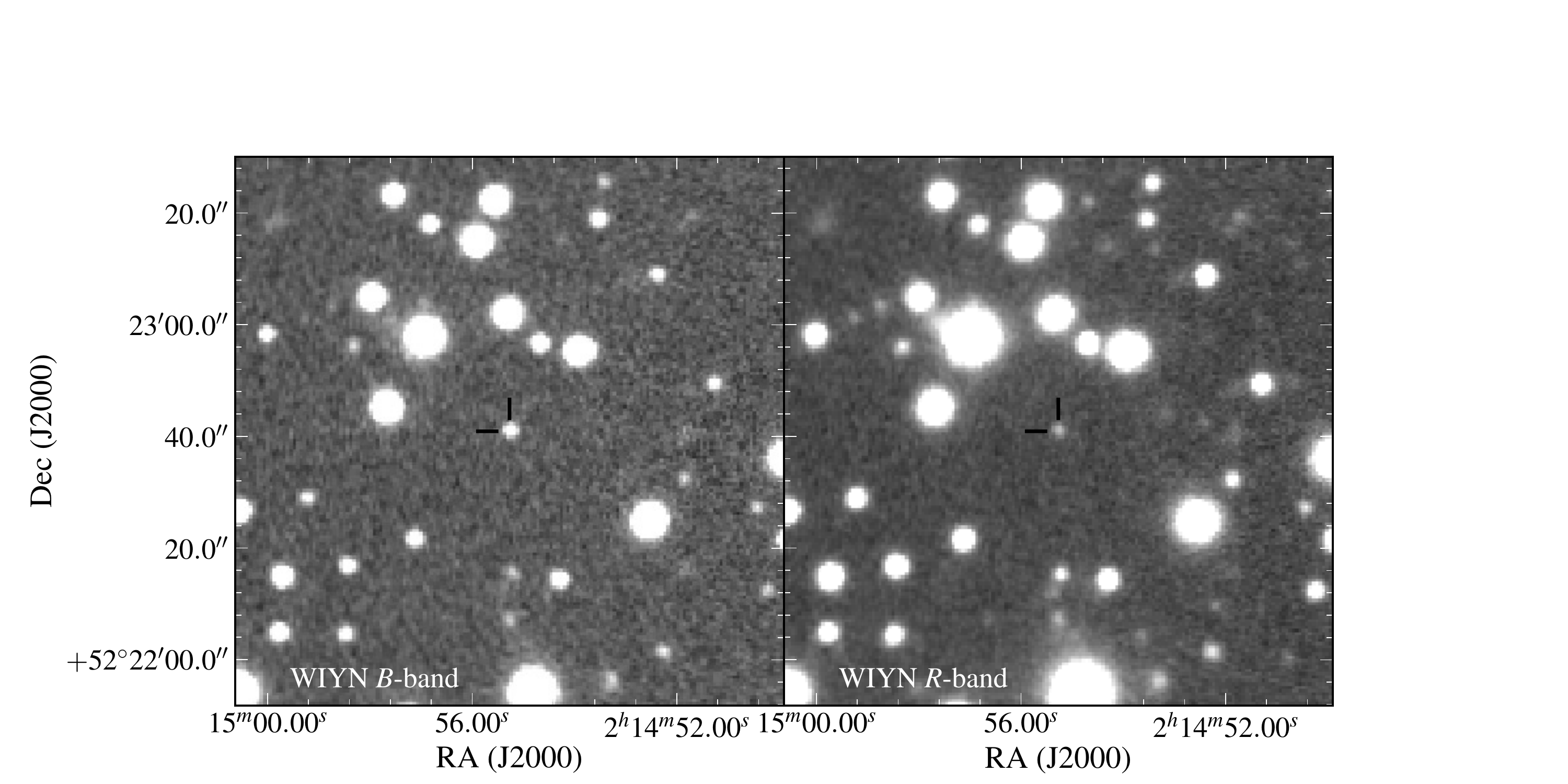}
\caption{Images of the field of PSR~J0214$+$5222 from the WIYN 0.9-m
  telescope.  The images are $1\farcm5$ on a side.  $B$-band is on the
  left, and $R$-band is on the right.  In each image we show ticks at
  the radio position of J0214$+$5222 which extend $2\arcsec$--$4\arcsec$
  from the pulsar. The astrometric uncertainty is far smaller, about
  $0\farcs2$.}
\label{fig:J0214_wiyn}
\end{figure}

PSR J0214$+$5222 (hereafter ``J0214'') is an intermediate-period pulsar with a spin period of 24.5~ms and
a DM of 22~$\mathrm{pc\,cm^{-3}}$. This pulsar is in a long-period, nearly circular orbit
of about 512~days with a projected semi-major axis of 175~lt-s. The minimum (where inclination = 90$^\circ$)
and median (assuming a distribution of random inclinations) companion masses
are 0.42~$\mathrm{M_\odot}$ and 0.49~$\mathrm{M_\odot}$, respectively. We compared the position to archival
optical catalogs and found a
possible counterpart at the position of J0214 in Second Generation Digitized Sky Survey
(DSS-2) images. In December of 2012, we observed the field containing J0214 with the
S2KB imager\footnote{\url{http://www.noao.edu/0.9m/observe/s2kb.html}} on the 0.9
meter WIYN\footnote{\url{http://www.noao.edu/0.9m/}}
telescope.  We took $6\times 10\,$min dithered exposures in the $B$
(2012~December~11) and $R$ (2012~December~12)
filters.  The data were reduced using normal procedures in
\texttt{IRAF}: we subtracted the bias using an overscan region and
separate bias frames, corrected the images using flatfields, and
combined the individual exposures. Astrometric solutions were derived
relative to the UCAC3 catalog \citep{2009yCat.1315....0Z}: we fit for
plate-scale, rotation, and center using 350 stars and found a solution
with an rms of $0\farcs1$ in each coordinate.  We determined a
photometric solution for each night using observations of the
\citet{2000PASP..112..925S}  standard field L98.  About 750 stars were
used for the solution (which we corrected to the airmass of the J0214
observations using the Kitt Peak extinction curves\footnote{See
  \url{http://www.noao.edu/kpno/manuals/dim/\#trans}.}), and we
estimate zero-point uncertainties of about $0.05\,$mag.  The images
are shown in Figure~\ref{fig:J0214_wiyn}.

We see an object extremely close to the position of the pulsar.  This
source is within $0\farcs16$ of the radio timing position, which we
compare with an estimated astrometric uncertainty of $0\farcs2$.  We
performed photometry using \texttt{sextractor}
\citep{ba96} and find $B=21.48\pm0.06$ and $R=21.72\pm0.09$ (including
zero-point uncertainties).  We estimate that there are $1.1\times
10^{-3}\,{\rm sources\,arcsec}^{-2}$ with $B<21.6$, so the chance of a
false association is roughly $10^{-4}$.  Because of this, we consider
it likely that we have detected the optical counterpart of the J0214
system, although we require spectroscopic confirmation to be certain.

It is already apparent from Figure~\ref{fig:J0214_wiyn} that the
counterpart is rather blue.  Most of the stars in the field have
$B-R\approx 1$, while the counterpart has $B-R=-0.13\pm0.11$.
However, with
only 2 bands we cannot uniquely determine the effective temperature of
the counterpart.  Instead we
estimate the extinction to be $A_\mathrm{V}\approx 0.4\,$mag at a distance of
1\,kpc (the DM distance) using \citet{dcllc03}.  With that we infer an
effective temperature of about 26,000\,K and a normalization of
$0.009\,R_\mathrm{\odot}/{\rm kpc}$.

This result is curious.  If the companion were a normal white dwarf,
the high temperature implies a young age, varying between 18\,Myr for
the lowest possible masses to 110\,Myr for 1\,$M_\mathrm{\odot}$ (we use the synthetic
photometry and evolutionary models from \citealt{tbg11} and
\citealt{bwd+11} for H and He atmospheres, respectively\footnote{Also see
  \url{http://www.astro.umontreal.ca/\textasciitilde bergeron/CoolingModels/}.}).  We have
ignored any possible heating of the companion by the pulsar, but due to the
long-period orbit this assumption seems reasonable.  This age is
much smaller than the characteristic age of J0214 (1.3 Gyr), and the radio
discovery of such a system during the short-lived time that the white dwarf
has such a high effective temperature seems rather fortuitous
(also see \citealt{2014arXiv1402.0407K}).

If the DM distance is close to correct, then the normalization is
inconsistent with a companion mass near the minimum.  Instead it would
require a companion mass of about $0.9\,M_\mathrm{\odot}$, impling that the
inclination angle is $\approx$33$^\circ$ (although such a massive
companion does relax the age constraint since the effective temperature
would be less than the one inferred above).  Instead, we might
expect J0214 to be like other wide binaries with companions near
0.4\,$M_\mathrm{\odot}$, and an evolutionary path like that of helium-core
white dwarfs \citep{tlk12}.{\bfref The companion to J0214 is directly on}
the \citet{ts99} curve if its mass is
near the minimum value (inclination near edge-on).  In that case the
distance would be closer to 2\,kpc: not an unreasonable disagreement
for a source at high Galactic latitude. However, we note that given the pulsar's
spin period of 24.5 ms and period derivative of 3$\times$10$^{-19}$
the pulsar is not fully recycled, so the validity of the~\citet{tlk12} scenario
is not clear.

A possible resolution to the age question which makes the distance
disagreement even worse would be if the companion were a subdwarf B
(sdB) star \citep{heber09}.  These objects, which can be slightly more
massive than the most massive helium-core white dwarfs (around
0.45\,$M_\mathrm{\odot}$), more naturally have effective temperatures above
20,000\,K due to core helium burning.  Except for the mass, the
evolution could be reasonably similar to that which produces a
helium-core WD.  However, sdB stars have radii of $\gtrsim
0.1\,M_\mathrm{\odot}$, which would require a distance of $>5\,$kpc.
Reconciling this with the DM distance may be difficult, but as directed
searches for pulsars around sdB start have so far not been successful~\citep{2011A&A...531A.125C},
it would make J0214 the first known PSR-sdB system.
An optical spectrum should
be able to distinguish between these scenarios given the very
different surface gravities ($10^7\,\mathrm{{cm\,s}^{-2}}$ for a $0.4\,M_\mathrm{\odot}$
WD, up to $10^9\,\mathrm{{cm\,s}^{-2}}$ for a massive WD, and
$<10^6\,\mathrm{{cm\,s}^{-2}}$ for a sdB star).

\subsection{PSR J0636$+$5129}\label{subsec:J0636}
Follow-up observations of PSR J0636$+$5129 (hereafter J0636) revealed that this bright, nearby MSP
is in a binary system with a very short orbital period of 96\,minutes (PSR J1311$-$3430 is the only
rotation-powered pulsar with a shorter orbital period, by 2 minutes) and the pulsar's projected
semi-major axis is 0.009 lt-s. Assuming a $1.4\;M_\mathrm{\odot}$ pulsar mass, this implies that the
minimum and median companion masses are about $7.4\,M_\mathrm{J}$ and $8.5\,M_\mathrm{J}$, respectively. The DM of this system
($11.1\;\mathrm{pc\,cm^{-3}}$) indicates a distance of only 0.5 kpc, using
the NE2001 model. Our timing up to this point
has enabled a measurement of a proper motion of 5.2$\mathrm{^{+1.3}_{-1.2}}$ $\mathrm{mas\;yr^{-1}}$ and an
annual parallax measurement of 4.7(5) mas. We corrected the parallax measurement
for Lutz-Kelker bias as described by~\cite{2010MNRAS.405..564V} and get a value of
4.27(56) mas. The parallax measurement implies a distance of 0.21$\mathrm{^{+0.03}_{-0.02}}$ kpc.
Combining our proper motion measurement with our our parallax distance of 0.21 kpc implies
a transverse velocity of 5~$\mathrm{km\;s^{-1}}$. {\bfref Even after accounting for the peculiar
velocity of the sun, it is clear that J0636 did not receive a significant kick at birth and is among
the lowest of all currently published values. From the
Galactic latitude for J0636 (18.6$^\circ$) and the parallax distance mentioned above, we calculate a distance
from the Galactic plane (z-height)
for this pulsar of only 0.06 kpc, which is consistent with its low transverse velocity.}

\begin{figure}[h!]
\centering
\includegraphics[width=0.5\textwidth]{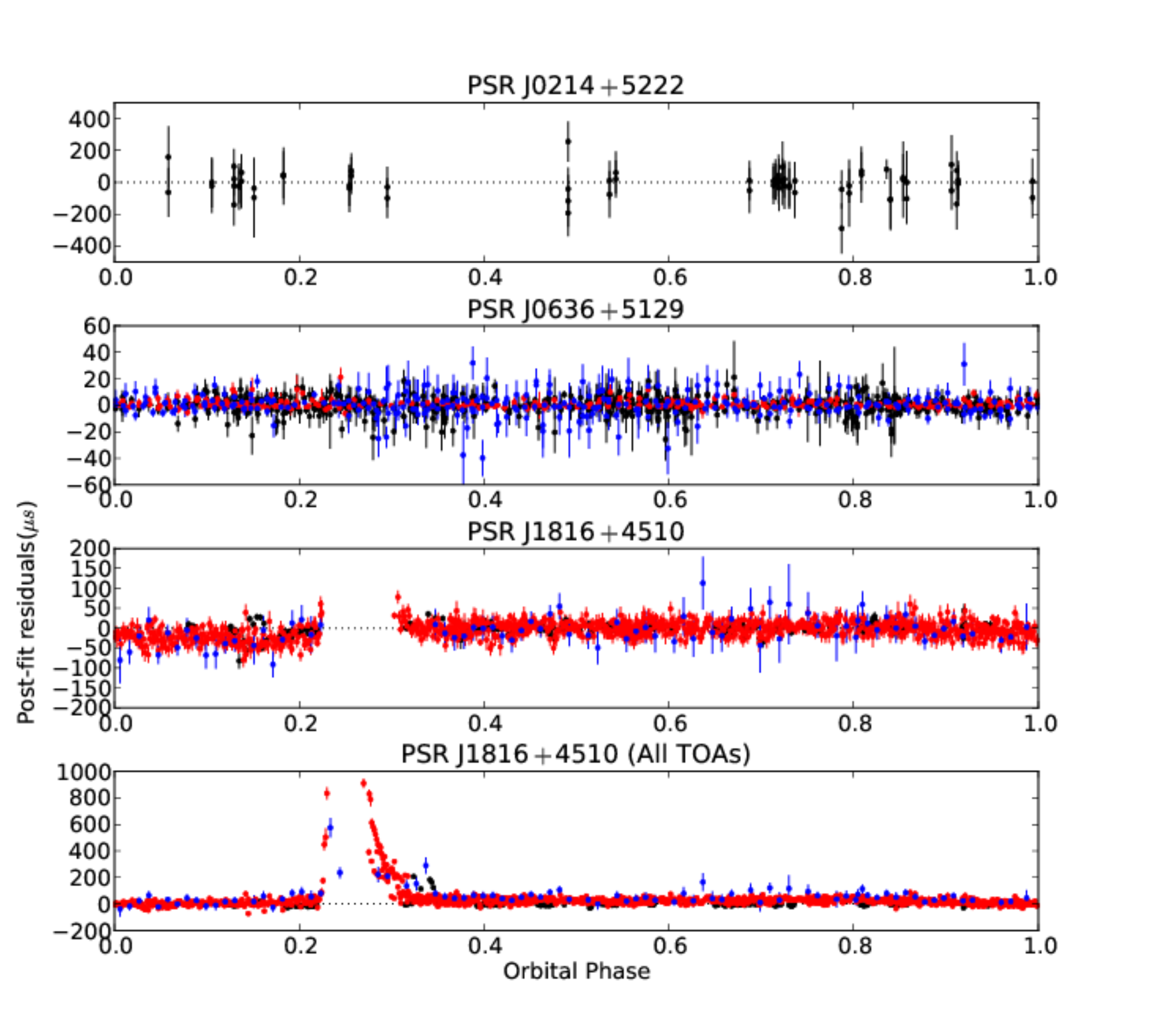}
\caption[Residuals vs Orbital Phase for GBNCC Binary MSPs]{TOA residuals vs orbital phase for PSRs J0214$+$5222, J0636$+$5129,
and J1816$+$4510. The third panel shows the residuals for PSR J1816$+$4510 with TOAs delayed due to extra material in the
system near eclipse removed. The bottom panel shows the residuals for J1816$+$4510 including TOAs near eclipse. Black
residuals are from 350-MHz observations, red were obtained from 820-MHz observations, and blue are from 1500-MHz observations.}
\label{fig:residsvorb}
\end{figure}

Initial assessment of this system based on the orbital period and companion mass would
suggest that this system is potentially related to the black widow class of pulsars. In
fact, the orbital period and companion mass of J0636 are very similar to the recently
discovered black widow system J1311$-$3430~\citep{2012Sci...338.1314P,2012ApJ...760L..36R,2013ApJ...763L..13R}.
However, unlike J1311$-$3430 and most other black widow systems, J0636 does not exhibit
eclipses and there are no signs of delays due to DM variations
caused by excess material in the system (see Fig.~\ref{fig:residsvorb}). Nor are there
any signs of changes in the orbital period over time. The fact that these effects, which
are commonly seen in black widow systems, are absent may indicate that
the system is instead related to the ``diamond planet'' system~\citep[J1719$-$1438;][]{2011Sci...333.1717B}
which consists of an MSP in a 140-minute orbit with a 1.2-$M_\mathrm{J}$ companion. We note that
there are some black widow systems~\citep[e.g.\;J0023$+$0923;][]{2011AIPC.1357...40H}
which do not show these effects in their radio timing, however they have been identified
as such systems based on optical follow-ups of their companion, which revealed periodic modulations in the
companion light curve consistant with irradiation by the pulsar wind~\citep{2013ApJ...769..108B}. Inspection of
archival data from First Generation Digitized Sky Survey (DSS-1), DSS-2, Two Micron All Sky
Survey (2MASS), and Wide-field Infrared Survey Explorer (WISE) revealed no optical
counterpart at the position of J0636. For comparison, if the very similar black widow
system J1311$-$3430 which has a DM estimated distance of 1.4~kpc were at the position
of J0636, its companion would have a magnitude of about 16 at maximum and 19.2 at
superior conjunction. 

An interesting aspect of the PSR J1719$-$1438 system is the large implied density of
the companion ($\gtrsim23 \mathrm{g\,cm^{-3}}$). This lower limit is derived from the
mean density-orbital period relation~\citep{frank_accretion_2002}, which does
not depend on the inclination of the system. 
Applying the same calculation to J0636, which has a tighter orbit and more massive
companion, gives roughly a factor of two denser companion $\gtrsim43 \mathrm{g\,cm^{-3}}$. 
The PSR J1719$-$1438 formation scenario presented by~\cite{2011Sci...333.1717B} is that the system
may have been an ultra compact X-ray binary (UCXB) system in which the companion was almost
completely destroyed by accretion. It has been postulated that this process could be aided by
some combination of a wind from the donor and/or evaporation by the pulsar wind and irradiation
through X-ray feedback~\citep{2012A&A...541A..22V,2012ApJ...753L..33B}.

J0636 may have undergone a similar history and therefore may be extremely useful in
characterizing such systems. We note that the orbital period and minimum companion mass
of this system agrees well with the relation in~\citet{2003ApJ...598.1217D}. \cite{2011Sci...333.1717B}
unsuccessfully attempted to detect
the companion of PSR J1719$-$1438 using the Keck 10-m telescope down to a limit of $R\sim24$.
However, J0636 is a better candidate for optical follow-up for multiple reasons. J0636
is nearby, it has high Galactic longitude ($163.9^\circ$) and latitude ($18.6^\circ$), and
the companion is more massive. Such observations would help to constrain the nature (radius,
temperature, and eventually composition and inclination) of the companion. This could then
be combined with studies of the expected evolutionary scenarios~\citep{2003ApJ...598.1217D,2012A&A...537A.104V}
and interior temperature to determine the history of this system and see if it is connected to UCXB systems.

In addition to being in an interesting binary system, J0636 is also a very fast-spinning, bright,
nearby MSP with a small duty cycle and therefore may be a useful source for PTAs. This source has
been released to the PTAs and is being tested for possible inclusion in their timing programs;
it is also a good candidate for VLBI observations due to its proximity and mean flux density at
1.4 GHz.

\subsection{PSR J0645$+$5158}\label{subsec:J0645}
PSR J0645$+$5158 is a bright, isolated MSP with a very small duty cycle (1-2\%)
that has already been added to the timing programs of PTAs. The
measured DM of 18.2 $\mathrm{pc\,cm^{-3}}$ implies a distance of about 0.7 kpc using
the NE2001 model for the Galactic electron density~\citep{2003astro.ph..1598C}. {\bfref PSR J0645$+$5158
shows signs of some scintillation, which is expected given the low DM, but typically has multiple
scintles within the observed bandwidth, so it is reliably detected. The observations centered at
820 MHz have provided the best timing results.} {\bfref The
current timing solution has a residual RMS of 0.51 $\mu s$, which } has allowed us to measure
the proper motion of {\bfref J0645$+$5158 to be} 7.56(25) $\mathrm{mas\;yr^{-1}}$ {\bfref and} an annual parallax of
1.4(3) mas. After correction for Lutz-Kelker bias, we get a parallax of 0.96(35) mas,
which results in a distance
measurement of 0.65$\mathrm{^{+0.20}_{-0.13}}$ kpc; this is in good agreement with the
distance obtained from the NE2001 model. Using the distance of 0.65 kpc, we measure a
transverse velocity for PSR J0645$+$5158 of 25 $\mathrm{km\;s^{-1}}$. {\bfref We note that
we have not accounted for proper motion due to the peculiar velocity of the sun
or due to differential Galactic rotation, both of which are
significant compared to the measured 25 $\mathrm{km\;s^{-1}}$.}

\subsection{PSR J1434$+$7257}\label{subsec:J1434}
PSR J1434$+$7257 is a partially recycled pulsar with a spin period of
41.7~ms. The timing solution indicates that it is an isolated pulsar, but its relatively low
period derivative indicated a characteristic age of 1.2~Gyr. The system is likely to
have been partially recycled and then disrupted. It may be the end result of a disrupted
neutron star system~\citep{1983ApJ...267..322H,1991PhR...203....1B}. {\bfref The Galactic
latitude of this system (42.2$^\circ$) combined with its DM-estimated distance of 0.7 kpc
gives an estimated z-height of 0.47 kpc. Though we have not yet detected a proper motion
for this system, we can place an upper limit of about 30 $\mathrm{mas\;yr^{-1}}$. Assuming the
DM estimated distance, this implies an upper limit on the transverse velocity for
PSR J1434$+$7257 of 100 $\mathrm{km\;s^{-1}}$.}

\subsection{PSR J1816$+$4510}\label{subsec:J1816}
PSR J1816$+$4510 (hereafter J1816) was the source we discovered by searching beams containing a \textit{Fermi} gamma-ray point
source. This MSP was discovered in a beam coincident with the \textit{Fermi} 1 year catalog~\citep{2010yCat..21880405A}
source 1FGL J1816.7$+$4509. Follow-up
observations were performed in the weeks after discovery showing that the MSP is in an eclipsing binary
system with an orbital period of 8.7 hours and a projected semi-major axis of 0.6 lt-s
with very little eccentricity. Assuming a pulsar mass of $1.4\;M_\mathrm{\odot}$, the companion has
a minimum mass of $0.16\;M_\mathrm{\odot}$ and a median predicted mass of $0.19\;M_\mathrm{\odot}$. The length of the
eclipses is approximately 7-10\% of the orbital period, which implies that the eclipsing region
has a minimum diameter of $1.3\;R_\mathrm{\odot}$. Assuming that the inclination of the system is
close to 90$^\circ$, the size of the Roche lobe of the companion is $R_\mathrm{L}\approx0.5\;R_\mathrm{\odot}$.
This indicates that if the
companion is filling its Roche lobe, then the diameter of the companion ($2R_\mathrm{L}\approx1.0\;R_\mathrm{\odot}$) is comparable
to the size of the eclipsing region. Delays in pulse TOAs are seen just before and after eclipses. If we assume
that these are due to an increased DM then we get an additional DM of up to
$0.14\;\mathrm{pc\;cm^{-3}}$. If we assume these electrons are constrained to a size comparable to
the Roche lobe of the companion, then the electron density in this region is up to $6\times10^{6}\;\mathrm{cm^{-3}}$.
The companion mass, eclipsing nature, and DM variations directly before and
after eclipse initially led us to believe this source is another redback system.

After a few months of follow-up observations, we were able to construct a phase connected solution which constrained
the position of J1816 to less than an arcsecond. Archival optical and infrared data were inspected
and revealed that a bright star is coincident with the pulsar's position~\citep{2012ApJ...753..174K}.
Additional optical observations have been performed and reveal that the star seen in the archival
data is the companion of J1816~\citep{2013ApJ...765..158K} and that the companion's estimated effective
temperature ($T_\mathrm{eff}$) is about 15,000 K. This temperature is
significantly greater than $T_\mathrm{eff}$ for any known redback system ($T_\mathrm{eff}\lesssim\mathrm{6,000}$ K).
Also, the spectroscopic study performed by~\cite{2013ApJ...765..158K} showed strong lines of helium and metals. J1816
seems to have several properties that make it distinct among the eclipsing pulsar systems.

J1816 was observed by the \textit{Chandra X-Ray Observatory} for 33.5~ksec, which is slightly longer than
one orbital period, using the ACIS-S detector on 2013 July 22. The data were analyzed using
CIAO{\bfref \footnote{http://cxc.cfa.harvard.edu/ciao/}} version 4.5 and Xspec{\bfref \footnote{http://heasarc.nasa.gov/xanadu/xspec/}} version 12.8.
A total of 17 photons were detected between 0.3 and 8 keV in
a $2^{\prime \prime}$ radius region centered on the timing position of the pulsar, with likely no more
than 2 being background photons. One of the photons had an energy of around 4~keV, and 4 in the 1.5-2~keV
range, suggesting some of the emission could be non-thermal, given the typical temperatures of thermal emission
from MSPs are in a fairly narrow range of ~0.15-0.2 keV~\citep[e.g.][]{2003A&A...398..639Z,2007ApJ...670..668B,2008ApJ...689..407B}.
However, it is likely that the thermal emission dominates as is the case for most MSPs. The total $0.3-8$~keV flux was
$4.3\pm1.1 \times 10^{-15} {\rm ergs}\,{\rm cm}^{-2}\,{\rm s}^{-1}$, assuming a model with $kT$=0.17~keV,
$nH$=0.02$\times 10^{20} {\rm cm}^{-2}$, and using a power-law index of 1.5. Given the high Galactic latitude and
distance estimated from either the DM or optical studies, the total X-ray absorption is
expected (eg. using the \texttt{HEASARC} nH tool based on~\cite{1990ARA&A..28..215D}) to be
$nH < 3.8\times 10^{20} {\rm cm}^{-2}$, which is quite low and so the unabsorbed flux is likely no more
than 15\% greater than the observed flux. Even given the larger distance estimated by~\cite{2013ApJ...765..158K}
of 4.5~kpc, the total X-ray luminosity is only $0.0002\dot E$, where $\dot E$ is the spindown luminosity given
in Table~\ref{tab:binaries}. This is typical for ordinary MSPs
suggesting there is very little X-ray emission arising from an intrabinary shock, unlike what is seen from
most redbacks{\bfref~\citep{2014AN....335..313R}}.

\begin{figure}[h!]
\centering
\includegraphics[width=0.5\textwidth]{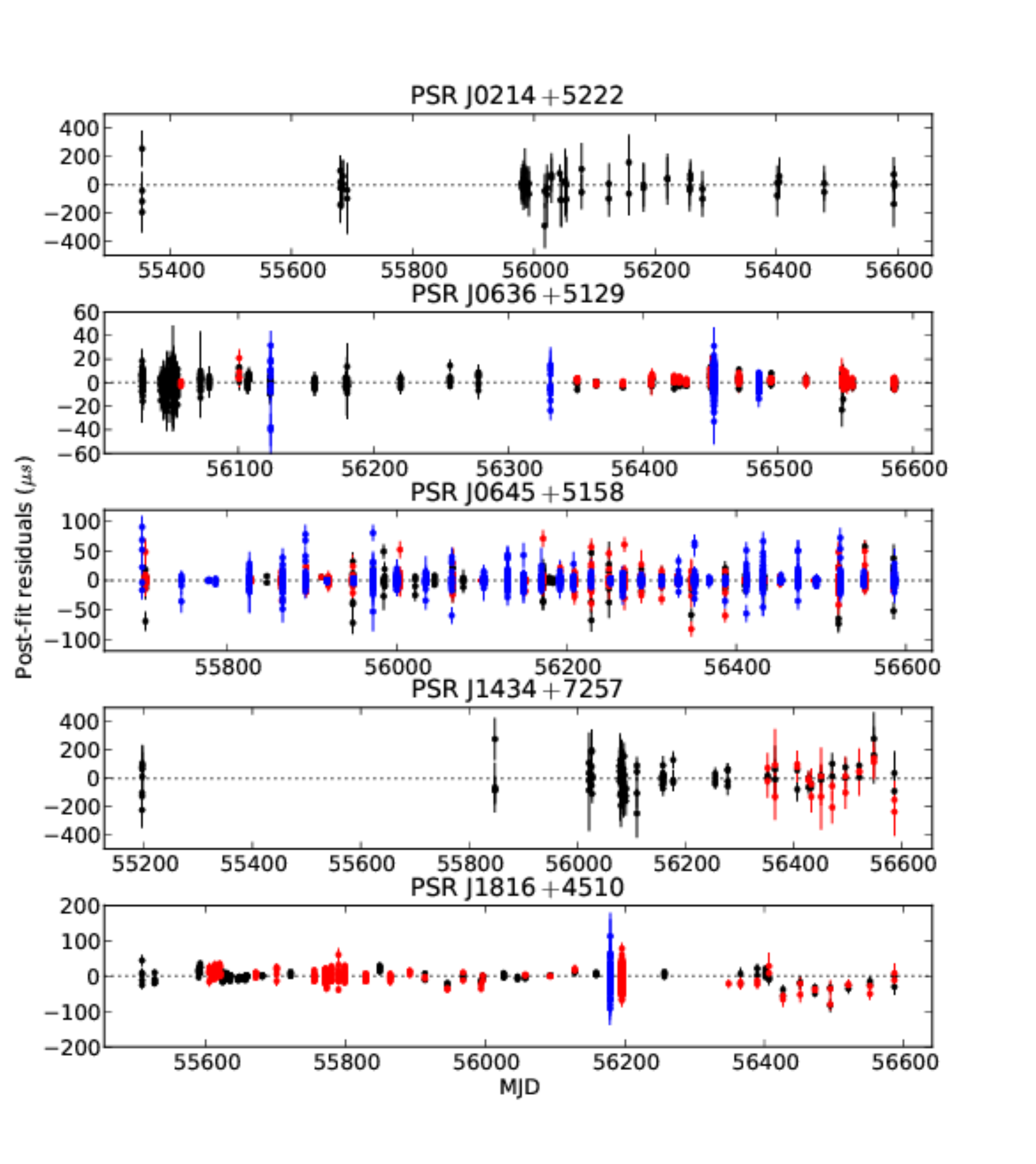}
\caption[Residuals for 5 GBNCC Pulsars]{Post-fit residuals in microseconds for five GBNCC pulsar discoveries. 
Black residuals were obtained from 350-MHz observations, red residuals are from 820-MHz observations, and blue
residuals were obtained from 1500-MHz observations.}
\label{fig:resids}
\end{figure}

\vspace{-2em}

\section{Conclusions}\label{sec:gbncc_conc}

The GBNCC is a 350-MHz pulsar and fast transient survey of the entire sky visible by the GBT.
Our sensitivity is about 2.5 times better than previous surveys for regions of the sky not
visible by the Arecibo Observatory. In the Arecibo visible range ($-2^\circ>\delta>38^\circ$),
we have comparable sensitivity to previous surveys for normal pulsars and
much better sensitivity to MSPs. {\bfref As of 2014 March 18,} we have completed observations for
about 35\% of the full survey and have analyzed about 75\% of the data recorded thus far. Using previously
known pulsars detected in our survey, we have compared expected survey sensitivity to actual sensitivity
and conclude that the results roughly agree. We have provided parameters for these re-detections (as well
as for our new discoveries), which will be useful for future pulsar population models.

To date, our survey has resulted in the discovery of 67 new radio pulsars, whose parameters were reported
here, as well as 7 new RRATs. Of the 67 new pulsars, 9 of them are MSPs. Comparison of the 11 MSPs detected
so far (9 new and 2 previously known) to the 30 MSPs expected for stage 1 from simulations suggest that
there are still more discoveries to be made. We note that about 20\% of the stage 1 data remains to
be processed {\bfref and that of the 80\% that has been processed, we have performed only a cursory examination
of a large fraction of the candidates. We estimate that the candidates have been examined at roughly the 50\% level,
so additional discoveries are likely in stage 1 data.} {\bfref A major issue for large scale pulsar surveys such
as the GBNCC is the difficulty in examining all of the candidates. We have used computer algorithms to generate
scores for each candidate and increased the number of people looking through the candidates by teaching high school
and undergraduate students to identify potential pulsar signals. These students have examined over 200,000
candidates and are responsible for the discovery of roughly 50 new pulsars in the GBNCC survey.}

We have reported discovery parameters for 62 of our discovered pulsars and
timing solutions for 5 more pulsars including two short-period binary MSPs (PSRs J0636$+$5129 and
J1816$+$4510), a long-period binary MSP (PSR J0214$+$5222), an isolated MSP (PSR J0645$+$5158), and an
isolated, partially recycled pulsar (PSR J1434$+$7257). PSR J0636$+$5129 has orbital and companion
mass parameters which are very similar to black widow pulsar systems, however it does not show
any evidence of the companion being ablated. It may instead be more similar to the diamond planet
pulsar systems. PSR J1816$+$4510 is an eclipsing binary system with an orbital period and companion
mass resembling redback pulsar systems, however optical follow-up has indicated that the companion
is quite unlike typical redback systems and may be young and that this system has recently completed
its spin-up phase. PSR J0214$+$5222 is a partially recycled MSP in a long-period orbit with a 0.4 $M_\mathrm{\odot}$
companion. Comparison to archival optical catalogs revealed an optical counterpart at the position
of J0214$+$5222 that is brighter than would be expected for typical pulsar systems with a white dwarf
companion. The companion may either be a very young white dwarf, or instead could be a subdwarf B star,
making this the first known PSR-sdB system. Two of our discoveries (PSR J0636$+$5129 and PSR J0645$+$5158)
are bright, low-DM MSPs which are likely to be useful in efforts to detect GWs using PTAs.

\section*{Acknowledgements}
The National Radio Astronomy Observatory is a facility of the National Science Foundation operated under
cooperative agreement by Associated Universities, Inc.
The optical results in subsection~\ref{subsec:J0214} is based upon data from the WIYN Observatory, which is a
joint facility of the University of Wisconsin-Madison, Indiana University, Yale University and the 
National Optical Astronomy Observatories. The authors acknowledge the Texas Advanced Computing
Center\footnote{\url{http://www.tacc.utexas.edu}} (TACC)
at The University of Texas at Austin for providing HPC resources that have contributed to the research results
reported within this paper. We also thank Compute Canada and the McGill Center for High Performance Computing and
Calcul Quebec (formerly CLUMEQ) for provision and maintenance of the Guillimin supercomputer and related resources.
This work was partially supported by the Chandra X-Ray Observatory Guest Observer program, SAO grant GO2-13056X.
KS was supported by the NSF through grants \#AST0545837 and \#AST0750913. Pulsar research at UBC is supported by an
NSERC Discovery Grant. Pulsar research at McGill is supported by an NSERC Discovery Grant and Accelerator Supplement,
the Canada Research Chair Program, CIFAR, FQRNT and the Lorne Trottier Chair in Astrophysics and Cosmology.
J.W.T.H. acknowledges funding from an NWO Vidi fellowship and ERC Starting Grant ``DRAGNET'' (337062).
This research has made use of the NASA/IPAC Infrared Science Archive, which is operated by the Jet Propulsion 
Laboratory, California Institute of Technology, under contract with the National Aeronautics and Space Administration.
{\bfref The authors of this manuscript thank Sarah~Battat and Zachary~Dionisopoulos for looking through pulsar
candidates during a summer program at McGill University.}
{\bfref The authors thank the referee, M.~Bailes, for his useful comments which greatly improved this manuscript.}

{\it Facilities:}  \facility{GBT}, \facility{WIYN:0.9m (S2KB)}, \facility{CXO}

\section*{Appendix}

\setcounter{table}{0}

Table~\ref{tab:redet} contains 64 observations of 55 previously known pulsars that have thus far been re-detected
in the GBNCC survey {\bfref within the 18' HWHM of the 350 MHz GBT beam}. The period and DM as detected by the GBNCC
survey are given, as well as the distance from the center of the beam in which the pulsar was detected, {\bfref and
the observed mean flux density at 350\,MHz (for a description, see sec~\ref{sec:gbnccredet}).}

\renewcommand{\thefootnote}{\alph{footnote}}
{\footnotesize
\begin{table*}[h]
\caption{Offset from GBNCC beam center and the estimated flux densities for 64 re-detections of 55 known pulsars detected in the GBNCC survey.}
\begin{center}
\begin{tabular}{ccc}
\hline
Pulsar & r & $\mathrm{S_{350}^{o}}$ \\
   & $^\circ$ & mJy \\
\hline
\hline
B0037+56 & 0.21 & 5.0 \\*
B0037+56 & 0.30 & 4.7 \\
\hline
B0059+65 & 0.24 & 29.7 \\
\hline
B0105+68 & 0.19 & 30.5 \\
\hline
B0114+58 & 0.30 & 43.4 \\
\hline
B0136+57 & 0.15 & 52.7 \\
\hline
B0138+59 & 0.25 & 64.9 \\
\hline
B0154+61 & 0.16 & 7.3 \\
\hline
J0215+6218 & 0.27 & 14.5 \\*
J0215+6218 & 0.28 & 17.2 \\
\hline
J0218+4232 & 0.26 & 57.7 \\*
J0218+4232 & 0.29 & 39.8 \\
\hline
B0226+70 & 0.24 & 40.0 \\
\hline
B0320+39 & 0.13 & 10.8 \\
\hline
J0341+5711 & 0.15\tablenotemark{b} & 364.7 \\
\hline
B0353+52 & 0.10 & 21.6 \\
\hline
B0355+54 & 0.21 & 46.8 \\
\hline
B0410+69 & 0.05 & 6.7 \\
\hline
B0450+55 & 0.15 & 49.9 \\
\hline
B0458+46 & 0.08 & 10.9 \\
\hline
B0609+37 & 0.30 & 8.8 \\
\hline
B0643+80 & 0.13 & 6.1 \\
\hline
B0655+64 & 0.20 & 10.1 \\
\hline
B0809+74 & 0.12 & 41.1 \\
\hline
B0841+80 & 0.06 & 9.0 \\
\hline
B1112+50 & 0.26 & 5.3 \\*
B1112+50 & 0.27 & 5.2 \\
\hline
B1322+83 & 0.25 & 22.4 \\*
B1322+83 & 0.26 & 14.1 \\
\hline
B1508+55 & 0.18 & 39.3 \\
\hline
J1518+4904 & 0.22 & 3.4 \\
\hline
B1839+56 & 0.16 & 57.9 \\
\hline
\end{tabular}
\quad
\quad
\begin{tabular}{ccc}
\hline
Pulsar & r & $\mathrm{S_{350}^{o}}$ \\
   & $^\circ$ & mJy \\
\hline
\hline
B1946+35 & 0.09 & 108.5 \\
\hline
B2000+40 & 0.30 & 4.7 \\
\hline
J2027+4557 & 0.06 & 14.2 \\
\hline
B2027+37 & 0.16 & 31.2 \\
\hline
B2035+36 & 0.20 & 12.6 \\
\hline
J2043+7045 & 0.28\tablenotemark{b} & 63.7 \\
\hline
B2045+56 & 0.17 & 18.5 \\
\hline
B2053+36 & 0.27 & 37.5 \\
\hline
B2106+44 & 0.28 & 37.0 \\
\hline
B2111+46 & 0.18 & 47.2 \\
\hline
J2217+5733 & 0.18 & 2.8 \\*
J2217+5733 & 0.18 & 4.6 \\
\hline
B2217+47 & 0.29 & 42.0 \\
\hline
J2222+5602 & 0.22\tablenotemark{b} & 21.1 \\
\hline
B2224+65 & 0.22 & 35.4 \\
\hline
J2229+6114 & 0.13 & 1.5 \\
\hline
B2227+61 & 0.17 & 24.9 \\
\hline
J2238+6021 & 0.16\tablenotemark{b} & 111.1 \\
\hline
B2241+69 & 0.19 & 6.3 \\*
B2241+69 & 0.19 & 6.6 \\
\hline
B2255+58 & 0.10 & 251.9 \\
\hline
J2302+6028 & 0.11 & 36.0 \\
\hline
J2302+4442 & 0.11 & 11.3 \\
\hline
B2306+55 & 0.25 & 122.2 \\*
B2306+55 & 0.26 & 53.0 \\
\hline
B2310+42 & 0.15 & 7.3 \\
\hline
B2319+60 & 0.17 & 303.2 \\
\hline
B2323+63 & 0.20 & 42.2 \\
\hline
B2324+60 & 0.11 & 26.9 \\*
B2324+60 & 0.11 & 27.7 \\
\hline
J2352+65 & 0.27\tablenotemark{b} & 26.5 \\
\hline
B2351+61 & 0.24 & 15.1 \\
\hline
\end{tabular}
\tablenotetext{1}{The position of these pulsars obtained from the ATNF database are only to the nearest 0.25$^{\circ}$.}
\label{tab:redet}
\end{center}
\end{table*} }
\renewcommand{\thefootnote}{\arabic{footnote}}

\newpage

\bibliography{gbncc}
\bibliographystyle{yahapj}

\end{document}